\documentclass[preprint,showpacs,preprintnumbers,amsmath,amssymb]{revtex4}
\usepackage{amsmath}
\newcommand\numberthis{\addtocounter{equation}{1}\tag{\theequation}}

\usepackage{graphicx}
\usepackage{pdfpages}
\usepackage{dcolumn}
\usepackage{bm}
\usepackage{amsmath} 

\newcommand{\qed}{\nobreak \ifvmode \relax \else
      \ifdim\lastskip<1.5em \hskip-\lastskip
      \hskip1.5em plus0em minus0.5em \fi \nobreak
      \vrule height0.75em width0.5em depth0.25em\fi}

\begin{document}

\preprint{}

\title{Formulas for Generalized Two-Qubit Separability Probabilities}
\author{Paul B. Slater}
 \email{slater@kitp.ucsb.edu}
\affiliation{%
Kavli Institute for Theoretical Physics, University of California, Santa Barbara, CA 93106-4030\\
}
\date{\today}
            
\begin{abstract}
To begin, we find certain formulas $Q(k,\alpha)= G_1^k(\alpha) G_2^k(\alpha)$, for $k = -1, 0, 1,\ldots 9$. These yield that part of the {\it total} separability probability, $P(k,\alpha)$,
for generalized (real, complex, quaternionic,\ldots) two-qubit states endowed with random induced measure, for which   the determinantal inequality $|\rho^{PT}| >|\rho|$ holds. Here $\rho$ denotes a $4 \times 4$ density matrix, obtained by tracing over the pure states in $4 \times (4 +k)$-dimensions, and $\rho^{PT}$, its partial transpose. Further, $\alpha$ is a Dyson-index-like parameter with $\alpha = 1$ for the standard (15-dimensional) convex set of (complex) two-qubit states.  For $k=0$, we obtain the previously reported Hilbert-Schmidt formulas, with
(the real case) $Q(0,\frac{1}{2}) = \frac{29}{128}$, (the standard complex case) $Q(0,1)=\frac{4}{33}$, and 
(the quaternionic case) $Q(0,2)= \frac{13}{323}$---the three  simply 
equalling $ P(0,\alpha)/2$. The factors $G_2^k(\alpha)$ are sums of polynomial-weighted generalized hypergeometric functions $_{p}F_{p-1}$, $p \geq 7$, 
all with argument $z=\frac{27}{64} =(\frac{3}{4})^3$. 
We find number-theoretic-based formulas for the upper ($u_{ik}$) and lower ($b_{ik}$) parameter sets of these functions and, then, equivalently express $G_2^k(\alpha)$ in terms of first-order difference equations. Applications of Zeilberger's algorithm yield  ``concise'' forms of $Q(-1,\alpha),Q(1,\alpha)$ and $Q(3,\alpha)$, parallel to the one obtained previously  ({\it J. Phys. A}, {\bf{46}} [2013], 445302) for $P(0,\alpha) =2 Q(0,\alpha)$.  For nonnegative half-integer and integer values of $\alpha$, $Q(k,\alpha)$ (as well as $P(k,\alpha)$) has descending   roots starting at $k=-\alpha-1$. Then, we (C. Dunkl and I) construct a remarkably compact (hypergeometric) form for $Q(k,\alpha)$ itself. The possibility of an analogous ``master'' formula for 
$P(k,\alpha)$ is, then, investigated, and a number of interesting results found.
\end{abstract}


\maketitle

\tableofcontents
\section{Introduction}
In a previous paper \cite{LatestCollaboration}, a family of  formulas was obtained for the (total) separability probabilities $P(k,\alpha)$ of generalized two-qubit states ($N=4$) endowed with Hilbert-Schmidt ($k=0$) \cite{szHS}, or more generally, random induced measure \cite{Induced,aubrun2}. In this regard, we note that the natural, rotationally invariant measure on the set of 
all pure states of a $N \times K$ composite system ($k=K-N$), induces a unique measure 
in the space of $N \times N$ mixed states \cite[eq. (3.6)]{Induced}.
Further, $\alpha$  serves as a Dyson-index-like parameter \cite{dyson,dumitriu}, assuming the values $\frac{1}{2}, 1, 2$ for the ($N=4$) two-rebit, (standard/complex) two-qubit, and two-quaterbit states, respectively.

The concept itself of a ``separability probability'', apparently first
(implicitly) introduced by {\.Z}yczkowski, Horodecki, Sanpera and Lewenstein in their much cited 1998 paper \cite{ZHSL}, entails computing the ratio of the volume--in terms of a given measure \cite{petzsudar}--of the separable quantum states to all quantum states.
Here, we first examine a certain component $Q(k,\alpha)$ of $P(k,\alpha)$. This informs us of that portion--equalling simply $P(k,\alpha)/2$  in the Hilbert-Schmidt ($k=0$) case \cite{WholeHalf}--for which the determinantal inequality 
$|\rho^{PT}| > |\rho|$ holds, with $\rho$ denoting a $4 \times 4$ density matrix and $\rho^{PT}$, its partial transpose. 
By consequence \cite{augusiak} of the Peres-Horodecki conditions \cite{asher,michal}, a necessary and sufficient condition for separability in this 
$4 \times 4$ setting is that 
$|\rho^{PT}|>0$. The nonnegativity condition $|\rho| \geq 0$ itself certainly holds, independently of any separability considerations. So, the total separability probability can clearly be expressed as the sum of that part for which $|\rho^{PT}| > |\rho|$ and that for 
which $|\rho| > |\rho^{PT}| \geq  0$. The former quantity will be the one of initial concern here, the ones the formulas $Q(k,\alpha)$ will directly yield. 

The complementary quantity, that for which 
$|\rho| > |\rho^{PT}| \geq  0$ can, in the most basic cases of interest, be 
readily obtained from the {\it total}
separability probability formulas $P(k,\alpha)$ reported in 
\cite{LatestCollaboration}, which took the form
\begin{equation} \label{HindawiFormula}
P(k,\alpha)  =1-F(k,\alpha),
\end{equation}
where for integral and half-integral $\alpha$,
\[
F\left(  k,\alpha\right)  =p_{\alpha}\left(  k\right)  G\left( k
,\alpha\right)  ,
\]
with
\[
G\left(  k,\alpha\right)  :=4^{k}\frac{\Gamma\left(  k+3\alpha+\frac{3}%
{2}\right)  \Gamma\left(  2k+5\alpha+2\right)  }{\Gamma\left(  \frac{1}%
{2}\right)  \Gamma\left(  3k+10\alpha+2\right)  }.
\]
Here, for integral $\alpha$, $p_{\alpha}\left(  k\right)  $ is a polynomial of degree $4\alpha-2$
with leading coefficient $\dfrac{2^{8\alpha+1}}{\left(  2\alpha-1\right)  !}.$

In \cite{LatestCollaboration}, certain $\alpha$-specific formulas ($\alpha = 1,2,\ldots,13$ and $\frac{1}{2}, \frac{3}{2},\frac{5}{2},\frac{7}{2}$) had been derived (and we have since continued the integral series to $\alpha=73$). Most notably  \cite[eq. (3)]{LatestCollaboration},
\begin{equation} \label{ComplexRule}
P(k,1)=1-\frac{3\ 4^{k+3} (2 k (k+7)+25) \Gamma \left(k+\frac{7}{2}\right) \Gamma (2k+9)}{\sqrt{\pi } \Gamma (3 k+13)}.
\end{equation}
Here $P(k,1)$ denotes the total separability probability of the (15-dimensional) standard, complex two-qubit systems endowed with the random induced measure for $k =K-4$.
Further, in the two-quater[nionic]bit setting \cite[eq. (4)]{LatestCollaboration},
\begin{equation} \label{QuaternionicRule}
P(k,2)=1-\frac{4^{k+6} (k (k (2 k (k+21)+355)+1452)+2430) \Gamma
   \left(k+\frac{13}{2}\right) \Gamma (2 k+15)}{3 \sqrt{\pi } \Gamma (3 k+22)}.
\end{equation}
Also, for the two-re[al]bit scenario \cite[eq. (5)]{LatestCollaboration},
\begin{equation} \label{RebitRule}
P(k,\frac{1}{2})=1-\frac{4^{k+1} (8 k+15) \Gamma (k+2) \Gamma \left(2
   k+\frac{9}{2}\right)}{\sqrt{\pi } \Gamma (3 k+7)}.
\end{equation}
Tables 1, 2 and 3 in \cite{LatestCollaboration} reported for $k =0,1,\ldots8$, the, in general, rather simple fractional separability probabilities $P(k,\alpha)$ yielded by these three formulas. 

By way of example, we first note that formula (\ref{ComplexRule}) yields
$P(1,1) = \frac{61}{143}$. Then, since we will find from our analyses below, that $Q(1,1)=\frac{45}{286}$, we can readily deduce that the corresponding 
(complementary) separability probability corresponding to the inequalities
$|\rho| > |\rho^{PT}| \geq  0$, for this 
$k=1, \alpha=1$ scenario is equal to $P(1,1)-Q(1,1)=\frac{7}{26} = \frac{61}{143}-\frac{45}{286}$. 

Let us further observe that for the Hilbert-Schmidt ($k=0$) case, strong evidence has been presented  \cite{WholeHalf} that for the two-rebit, two-qubit and two-quaterbit cases, the apparent total separability probabilities $P(0,\alpha)$ of $\frac{29}{64}, \frac{8}{33}$ and $\frac{26}{323}$, respectively, are equally divided between the two forms of determinantal inequalities (cf. \cite{sbz}. Lovas and Andai have recently formally proven this two-rebit result and presented an integral formula they hope to similarly yield the two-qubit proportion \cite{lovasandai}. (These ``half-probabilities'', remarkably, are also the corresponding separability probabilities of the {\it minimally degenerate} states \cite{sbz}, those for which $\rho$ has a zero eigenvalue.) For $k > 0$, however, our analyses will indicate that  equal splitting is not, in fact, the case. Greater separability probability is associated with the $|\rho| > |\rho^{PT}| \geq  0$ inequality than  $|\rho^{PT}| >|\rho|$. Thus, in the $k=1,\alpha=1$ instance just discussed, we do have
$\frac{7}{26} >\frac{45}{286}$. (On the other hand, if $k=-1$, then necessarily $|\rho|=0$, so all the total
separability probability $P(-1,\alpha)$ must, it is clear, be assigned to the 
$|\rho^{PT}]>|\rho|$ component. That is, $Q(-1,\alpha) = P(-1,\alpha)$.) 
Observations of this nature should help in the 
further understanding of 
the intricate geometry of the generalized two-qubit states endowed with random induced measure (cf. \cite{Gamel}).
\section{Procedures}
\subsection{Previous Analyses}
To obtain the new formulas $Q(k,\alpha)$ to be presented here 
for the separability probability amounts for 
which $|\rho^{PT}| > |\rho|$ holds, we first employed--as in our prior studies 
\cite{MomentBased,slaterJModPhys,WholeHalf,LatestCollaboration}--the Legendre-polynomial-based probability density approximation (Mathematica-implemented) algorithm of Provost 
\cite{Provost} (cf. \cite{askey1982hausdorff}). In this regard, we utilized the previously-obtained determinantal moment 
formula \cite[eq. (6)]{LatestCollaboration} \cite[sec. II]{WholeHalf}  (cf. \cite{Beran})
\begin{align*}
\left\langle \left\vert \rho\right\vert
^{k}\left(  \left\vert \rho^{PT}\right\vert -\left\vert \rho\right\vert
\right)  ^{n}\right\rangle /\left\langle \left\vert \rho\right\vert
^{k}\right\rangle   &  =
\left(  -1\right)  ^{n}\frac{\left(  \alpha\right)  _{n}\left(  \alpha
+\frac{1}{2}\right)  _{n}\left(  n+2k+2+5\alpha\right)  _{n}}{2^{4n}\left(
k+3\alpha+\frac{3}{2}\right)  _{n}\left(  2k+6\alpha+\frac{5}{2}\right)
_{2n}}\\
& \times~_{4}F_{3}\left(
\genfrac{}{}{0pt}{}{-\frac{n}{2},\frac{1-n}{2},k+1+\alpha,k+1+2\alpha
}{1-n-\alpha,\frac{1}{2}-n-\alpha,n+2k+2+5\alpha}%
;1\right) 
\end{align*}
(where the variable $k$ has the same sense as indicated above, in equalling $K-4$, and the bracket notation indicates averaging with respect to the random induced measure).
Here, $\left\langle \left\vert \rho\right\vert
^{k}\right\rangle =\frac{\left(  1\right)  _{k}\left(  \frac{3}{2}\right)
_{k}\left(  2\right)  _{k}\left(  \frac{5}{2}\right)  _{k}}{\left(  10\right)
_{4k}}$, where the Pochhammer (rising factorial) notation is employed.

On the other hand in \cite{LatestCollaboration}, a second companion moment formula \cite[sec. X.D.6]{MomentBased}
\begin{gather*} \label{nequalzero}
\left\langle \left\vert \rho^{PT}\right\vert ^{n}\right\rangle =\frac
{n!\left(  \alpha+1\right)  _{n}\left(  2\alpha+1\right)  _{n}}{2^{6n}\left(
3\alpha+\frac{3}{2}\right)  _{n}\left(  6\alpha+\frac{5}{2}\right)  _{2n}}\\
+\frac{\left(  -2n-1-5\alpha\right)  _{n}\left(  \alpha\right)  _{n}\left(
\alpha+\frac{1}{2}\right)  _{n}}{2^{4n}\left(  3\alpha+\frac{3}{2}\right)
_{n}\left(  6\alpha+\frac{5}{2}\right)  _{2n}}~_{5}F_{4}\left(
\genfrac{}{}{0pt}{}{-\frac{n-2}{2},-\frac{n-1}{2},-n,\alpha+1,2\alpha
+1}{1-n,n+2+5\alpha,1-n-\alpha,\frac{1}{2}-n-\alpha}%
;1\right)  \numberthis \label{MomentFormula}
\end{gather*}
had been utilized for density-approximation purposes with the routine of Provost, with the objective of finding the {\it total} separability probabilities $P(k,\alpha)$, associated with the Peres-Horodecki-based inequality
$|\rho^{PT}|>0$. (These moment formulas had been developed in \cite{MomentBased}, based on calculations solely for the two-rebit [$\alpha=\frac{1}{2}$] and two-qubit [$\alpha=1$] cases. However, they do appear, as well, remarkably, to apply to the two-quater[nionic]bit [$\alpha =2$] case, as reported by Fei and Joynt in a highly computationally intensive Monte Carlo study \cite{FeiJoynt}. No explicit formal extension of the Peres-Horodecki positive-partial-transposition separability conditions \cite{asher,michal} to two-quaterbit systems seems to have been developed, however [cf. \cite{carl,aslaksen,asher2}]. The value
$\alpha=4$ corresponds, presumably it would seem, to an {\it octonionic} setting 
\cite{najarbashi2016two,2016arXiv161008081F}.) 
\subsection{Present Analyses}
Here,  contrastingly (``dually'') with respect to the approach indicated in \cite{LatestCollaboration}, we will find $k$-specific formulas 
($k=-1,0,1,\ldots,9$) as a function of $\alpha$, that is $Q(k,\alpha)$, for the indicated one ($|\rho^{PT}| > |\rho|$) of the two component determinantal inequality parts of $|\rho^{PT}|>0$.
We utilized an exceptionally large number (15,801) number of the first set of moments 
above in the routine of Provost \cite{Provost}, helping to reveal--to extraordinarily
high accuracy--the rational values that the corresponding desired (partial) separability  probabilities $Q(k,\alpha)$ strongly appear to assume.
Sequences ($\alpha =1, 2,\ldots,30,\ldots$) of such rational values, then, served as input to the  FindSequenceFunction command of Mathematica, which then yielded  the initial set of $k$-specific (hypergeometric-based) formulas for $Q(k,\alpha)$. (This apparently quite powerful [but ``black-box'']  command of which we have previously and will now make copious use, has been described as attempting ``to find a simple function that yields the sequence  when given successive integer arguments''. It can, it seems, succeed too, at times, for rational-valued inputs, and perhaps even ones of a symbolic nature.) We, then, decompose $Q(k,\alpha)$ into the product form $G_1^k(\alpha) G_2^k(\alpha)$
\section{Common features of the $k$-specific formulas $Q(k,\alpha)$} \label{Common}
For each $k = -1, 0, 1,\ldots,9$, the FindSequenceFunction command yielded what we can
consider as a large, rather cumbersome (several-page) formula, which we denote by 
$Q(k,\alpha)$. These expressions, in fact, faithfully reproduce the rational-valued (separability probability) sequences that served as input. This fidelity is indicated by numerical calculations to apparently arbitrarily high accuracy (hundreds of digits). (The difference equation results below [sec.~\ref{Equivalent}] will provide a basis for our observation as to the rational-valuedness [fractional nature] of these separability probabilities.)

\begin{figure}
\includegraphics{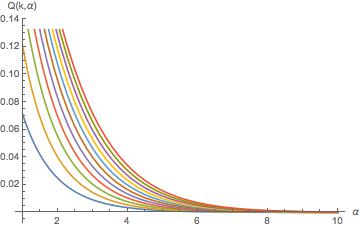}
\caption{\label{fig:Raw}Plots of the separability probability formulas 
$Q(k,\alpha)$ over the range
 $\alpha \in [1,10]$, for $k=-1,\ldots,9$. For fixed $\alpha$, we have $Q(k_1,\alpha) > Q(k_2,\alpha)$, if $k_1>k_2$.}
\end{figure}

\begin{figure}
\includegraphics{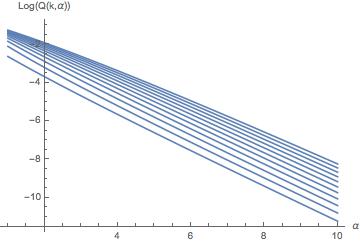}
\caption{\label{fig:Log}Plots of $\log{Q(k,\alpha)}$ over the range
 $\alpha \in [1,10]$, for $k=-1,\ldots,9$. For fixed $\alpha$, $\log{Q(k_1,\alpha)} > \log{Q(k_2,\alpha)}$, if $k_1>k_2$.}
\end{figure}

In Fig.~\ref{fig:Raw}, we show plots of the formulas $Q(k,\alpha)$ obtained over the range
 $\alpha \in [1,10]$, for $k=-1,\ldots,9$. For fixed $\alpha$, we have $Q(k_1,\alpha) > Q(k_2,\alpha)$, if $k_1>k_2$. In Fig.~\ref{fig:Log}, we show a companion plot, exhibiting strongly log-linear-like behavior, for $\log{Q(k,\alpha)}$.
\subsection{Distinguished $_{7}F_{6}$ function with 2 as an upper parameter in  $Q(k,\alpha)$} \label{First7F6}
In each of the eleven $k$-specific formulas $Q(k,\alpha)$ obtained, there is a distinguished $_{7}F_{6}$ generalized hypergeometric function, with
the (``omnipresent'', we will find) argument of  $z=\frac{27}{64} =(\frac{3}{4})^3$ (cf. \cite{guillera} \cite[Ex. 8.6, p. 159]{koepf}), 
having  2 as one of the seven upper parameters (cf. \cite{slaterJModPhys}). 
\subsubsection{The six lower parameters} \label{LowerParameters}
The lower (bottom) six parameters $b_{ik}$, $i=1,\ldots,6$, of the $_7F_6$ function 
conform for all eleven cases to the simple linear rule,
\begin{equation} \label{bottom}
\{b_{1k},b_{2k},b_{3k},b_{4k},b_{5k},b_{6k}\}=
\end{equation}
\begin{displaymath}
\left\{\alpha +\frac{2 k}{5}+\frac{23}{10},\alpha +\frac{2 k}{5}+\frac{5}{2},\alpha
   +\frac{2 k}{5}+\frac{27}{10},\alpha +\frac{2 k}{5}+\frac{29}{10},\alpha +\frac{2
   k}{5}+\frac{31}{10},\alpha +k+3\right\}.
\end{displaymath}
The six entries sum to $6 \alpha +3 k+\frac{33}{2}$.
\subsubsection{The six upper parameters} \label{UpperParameters}
The six  upper parameters (aside from the seventh $k$-invariant constant of 2 already 
indicated), $\{u_{1k},u_{2k},u_{3k},u_{4k},u_{5k},u_{6k}\}$,  can be broken into one set of two (the numerical parts summing to integers), incorporating consecutive fractions having 6's in their denominators, 
and one set of four (the numerical parts also summing to integers), incorporating consecutive fractions having 5's in their denominators. 
For the set of two, the smaller of the two upper entries abides by the rule
\begin{equation} \label{upper1}
u_{1k} =\alpha+ \frac{1}{6} \left(4 \left\lfloor \frac{k}{3}\right\rfloor +2 \left\lfloor
   \frac{k+1}{3}\right\rfloor +11\right),
\end{equation}
where the (integer-valued) floor function is employed, while 
the larger entry is given by
\begin{equation} \label{upper2}
u_{2k}= \alpha+\frac{1}{6} \left(2 \left\lfloor \frac{k}{3}\right\rfloor +4 \left\lfloor
   \frac{k+1}{3}\right\rfloor +13\right).
\end{equation}
For integral values of $k$, the same values of $u_{i1}$ and $u_{i2}$ are yielded by the interpolating functions,
\begin{displaymath}
\alpha+\frac{1}{18} \left(6 k+2 \cos \left(\frac{2 \pi  k}{3}\right)+2 \cos \left(\frac{4 \pi 
   k}{3}\right)+29\right),
\end{displaymath}
and
\begin{displaymath}
\alpha+ \frac{1}{18} \left(6 k-\sqrt{3} \sin \left(\frac{2 \pi  k}{3}\right)+\sqrt{3} \sin
   \left(\frac{4 \pi  k}{3}\right)+\cos \left(\frac{2 \pi  k}{3}\right)+\cos
   \left(\frac{4 \pi  k}{3}\right)+37\right),
\end{displaymath}
respectively.

For $k = 1$, for illustrative purposes, application of the two rules yields 
$\left\{\alpha +\frac{11}{6},\alpha +\frac{13}{6}\right\}$, and 
for $k=5$, we have $\left\{\alpha +\frac{19}{6},\alpha +\frac{23}{6}\right\}$. (We have noted that $u_{1k}+u_{2k} -2 \alpha =\left\lfloor \frac{k}{3}\right\rfloor +\left\lfloor \frac{k+1}{3}\right\rfloor +4$ is an integer. The sequence of these integers--for arbitrary integer or half-integer values of $\alpha$--is found
 in the On-Line Encyclopedia of Integer Sequences [https://oeis.org/ol.html] as A004523 
[``Two even followed by one odd"] and as
A232007 [``Maximal number of moves needed to reach every square by a knight from a fixed position on an n X n chessboard, or -1 if it is not possible to reach every square"].)

For the complementary set of four upper parameters of the $_7F_6$ function, 
the entries in order of increasing magnitude are expressible as 
\begin{equation} \label{upper3}
u_{3k}=\alpha+\frac{1}{5} \left(3 \left\lfloor \frac{k-4}{5}\right\rfloor +2 \left\lfloor
   \frac{k-3}{5}\right\rfloor +2 \left\lfloor \frac{k-2}{5}\right\rfloor +3 \left\lfloor
   \frac{k-1}{5}\right\rfloor +16\right),
\end{equation}
\begin{displaymath}
u_{4k}=\alpha +\frac{1}{5} \left(3 \left\lfloor \frac{k-4}{5}\right\rfloor +2 \left\lfloor
   \frac{k-3}{5}\right\rfloor +\left\lfloor \frac{k-2}{5}\right\rfloor +4 \left\lfloor
   \frac{k-1}{5}\right\rfloor +17\right),
\end{displaymath}
\begin{displaymath}
u_{5k}=\alpha +\frac{1}{5} \left(2 \left\lfloor \frac{k-4}{5}\right\rfloor +3 \left\lfloor
   \frac{k-3}{5}\right\rfloor +\left\lfloor \frac{k-2}{5}\right\rfloor +4 \left\lfloor
   \frac{k-1}{5}\right\rfloor +18\right),
\end{displaymath}
and
\begin{displaymath}
u_{6k}=\alpha +\frac{1}{5} \left(2 \left\lfloor \frac{k-4}{5}\right\rfloor +3 \left\lfloor
   \frac{k-3}{5}\right\rfloor +\left\lfloor \frac{k-2}{5}\right\rfloor +4 \left\lfloor
   \frac{k-1}{5}\right\rfloor +19\right).
\end{displaymath}
For $k = 1$, for illustrative purposes, application of these four rules yields $\left\{\alpha +\frac{9}{5},\alpha +\frac{11}{5},\alpha +\frac{12}{5},\alpha
   +\frac{13}{5}\right\}$, and 
for $k=5$, we have $\left\{\alpha +\frac{16}{5},\alpha +\frac{17}{5},\alpha +\frac{18}{5},\alpha
   +\frac{19}{5}\right\}$. For arbitrary $k$, the sum of the four terms under discussion minus $4 \alpha$
is an integer, namely, $2 \left\lfloor \frac{k-4}{5}\right\rfloor +2 \left\lfloor \frac{k-3}{5}\right\rfloor
   +\left\lfloor \frac{k-2}{5}\right\rfloor +3 \left\lfloor \frac{k-1}{5}\right\rfloor
   +14$.
Further, let us note that for integral values of $k$, $u_{3k}$ has values
\begin{center}
$\let\left\relax\let\right\relax
\frac{1}{250} \left(50 (2 k+7)+\sqrt{50-10 \sqrt{5}} \left(-3 \sin \left(\frac{2}{5} \pi 
   (1-2 k)\right)+2 \sin \left(\frac{4 \pi  k}{5}\right)-2 \sin \left(\frac{6 \pi 
   k}{5}\right)-3 \sin \left(\frac{1}{5} (\pi -6 \pi  k)\right)+2 \sin \left(\frac{1}{5}
   (\pi -4 \pi  k)\right)-3 \left(\sin \left(\frac{2}{5} (3 \pi  k+\pi )\right)+\sin
   \left(\frac{1}{5} (4 \pi  k+\pi )\right)\right)+2 \sin \left(\frac{1}{5} (6 \pi  k+\pi
   )\right)\right)+\sqrt{10 \left(5+\sqrt{5}\right)} \left(-2 \sin \left(\frac{2}{5} \pi 
   (1-4 k)\right)-2 \sin \left(\frac{2 \pi  k}{5}\right)+2 \sin \left(\frac{8 \pi 
   k}{5}\right)-2 \sin \left(\frac{2}{5} \pi  (k+1)\right)-3 \left(\sin \left(\frac{1}{5}
   (\pi -2 \pi  k)\right)-\sin \left(\frac{2}{5} (4 \pi  k+\pi )\right)+\sin
   \left(\frac{1}{5} (8 \pi  k+\pi )\right)\right)+3 \cos \left(\frac{1}{10} (4 \pi 
   k+\pi )\right)\right)\right)$.
\end{center}
\subsection{Distinguished $_{7}F_{6}$ function with 1 as an upper parameter in $Q(k,\alpha)$} \label{Isolated}
Each $k$-specific formula $Q(k,\alpha)$ we have found 
also incorporates a second 
$_{7}F_{6}$ function (again with argument $z=\frac{27}{64}$, which is, to repeat, invariably the case throughout this paper), having all its thirteen parameters simply equalling 1 less than those in the function just described. 
(A basic transformation exists 
[consulting the HYP manual of C. Krattenthaler, available at www.mat.univie.ac.at], allowing one to convert the thirteen [twelve  $\alpha$-dependent parameters, plus 1]  of this $_7F_6$ function [that is, add 1 to each of them] to those thirteen of the first distinguished
$_7F_6$ previously described, plus other terms.)

\subsection{The $m_k$ remaining $_{p}F_{p-1}$ functions, $p=8,\ldots,8+m_k-2$, in $Q(k,\alpha)$.}
Now, in addition to the two distinguished $_{7}F_{6}$ functions just presented, there are $m_k$ more hypergeometric functions $_pF_{p-1}$, $p>7$, for each $k$, where
\begin{equation} \label{numbers}
\{m_{-1},m_0,m_1,m_2,m_3,m_4,m_5,m_6,m_7,m_8,m_9\}=\{3, 5, 5, 6, 6, 7, 9, 8, 10, 10, 10\}.
\end{equation}
Each of these additional functions possesses, to begin with, 
the same seven upper parameters (that is, 2, plus those six indicated in 
(\ref{upper1}), (\ref{upper2}) and (\ref{upper3})) and the same six lower parameters (\ref{bottom}), as in the first $_{7}F_{6}$ function detailed above 
(sec.~\ref{First7F6}). Then, the seven upper parameters are supplemented by from 1 to 
$m_k$ 2's, and the six lower parameters supplemented by from 1 to   $m_k$ 1's.
\subsection{Large $\alpha$-free terms collapsing to 0}
We now point out a rather remarkable property of the formulas for $Q(k,\alpha)$ yielded by the  FindSequenceFunction command. If we isolate those (often quite bulky) terms that do not involve any of the $m_k+2$ hypergeometric functions for each $k$ already described above, we find (to hundreds of digits of accuracy) that they collapse to zero. These terms, typically, do contain hypergeometric functions similar in nature to those described above, but with the crucial difference that the Dyson-index-like parameter $\alpha$ does {\it not}
occur among their upper and lower parameters. Thus, we are left, after this nullification of terms, with formulas 
$Q(k,\alpha)$ that are simply sums of  $m_k+2$ polynomial-weighted $_pF_{p-1}$ functions (of $\alpha$), with $p=7,7, 8,\ldots,7+m_k$.
\subsection{Summary}
To reiterate, for each $k$, our formulas for $Q(k,\alpha)$, all contain a single function of the form
\begin{equation}
\, _7F_6\left(2,u_k,u_{2 k},u_{3 k},u_{4 k},u_{5 k},u_{6 k}; \newline b_k,b_{2 k},b_{3 k},b_{4
   k},b_{5 k},b_{6 k};\frac{27}{64}\right).
\end{equation}
There is another distinguished single $_7F_6$ function, with all its thirteen parameters being one less.
Also there are $m_k$ additional functions, $i=1,\ldots,m_k$,
\begin{displaymath}
\, _{7+i}F_{6+i}\left(2,2,\ldots,u_k,u_{2 k},u_{3 k},u_{4 k},u_{5 k},u_{6 k};1,\ldots,b_k,b_{2 k},b_{3 k},b_{4
   k},b_{5 k},b_{6 k};\frac{27}{64}\right),
\end{displaymath}
with the number of upper 2's running from 2 to $m_k+1$ and the number of lower 1's, 
simultaneously running from 1 to $m_k$.

\section{Decomposition of $Q(k,\alpha)$  into the product $G_1^k(\alpha) G_2^k(\alpha)$}
The formulas for  $Q(k,\alpha)$ that we have obtained can all be written--we have found--in the product form $G_1^k(\alpha) G_2^k(\alpha)$. The  $G_2^k(\alpha)$ factor involves the summation of the hypergeometric functions $_{p}F_{p-1}$ indicated above, each such function weighted by a polynomial in $\alpha$, the degrees of the weighting polynomials diminishing as $p$ increases. Let us first discuss the other (hypergeometric-free) factor $G_1^k(\alpha)$, 
involving ratios of products of Pochhammer symbols.
\subsection{Hypergeometric-function-independent factor $G_1^k(\alpha)$} \label{G1sec}
Some supplementary computations (involving an independent use of the \newline FindSequenceFunction command) indicated that these
(hypergeometric-free) factors can be written quite concisely, in terms of the upper and lower parameter sets, setting $U_{ik}=u_{ik}-\alpha, B_{ik} =b_{ik}+1-\alpha$, as
\begin{equation} \label{G1}
G_1^k(\alpha)= (\frac{27}{64})^{\alpha-1} \frac{\left(U_{1k}\right)_{\alpha -1} \left(U_{2k}\right)_{\alpha -1} \left(U_{3k}\right)_{\alpha
   -1} \left(U_{4k}\right)_{\alpha -1} \left(U_{5k}\right)_{\alpha -1} \left(U_{6k}\right)_{\alpha
   -1}}{\left(B_{1k}\right)_{\alpha -1} \left(B_{2k}\right)_{\alpha -1} \left(B_{3k}\right)_{\alpha
   -1} \left(B_{4k}\right)_{\alpha -1} \left(B_{5k}\right)_{\alpha -1} \left(B_{6k}\right)_{\alpha
   -1}},
\end{equation}
where the Pochhammer symbol (rising factorial) is employed. We note that, remarkably,
$G_1^k(1)=1$--further apparent indication of the special/privileged status of the standard (complex, $\alpha=1$) two-qubit states.
\subsection{Hypergeometric-function-dependent factor $G_1^k(\alpha)$}
\subsubsection{Canonical form}
In App.~\ref{fig:kminus1-kplus2}, for $k=-1,0,1,2$, we show the ``canonical form'' we have 
developed for the factors $G_2^k(\alpha)$ (cf. 
\cite[Fig. 3]{slaterJModPhys}), the component hypergeometric  parts of which we have discussed in sec.~\ref{Common}.
\section{Difference equation formulas for $G_2^k(\alpha)$} \label{Equivalent}
It further appears that all the $G_2^k(\alpha)$ factors ($k = -1,0,1,\ldots,9$)
(App.~\ref{DiffEqs}) can be equivalently written as functions that satisfy first-order difference (recurrence) equations of the 
form
\begin{equation} \label{recurrence}
p_0^k(\alpha) +p_1^k(\alpha) G_2^k(\alpha) +p_2^k(\alpha) G_2^k({1+\alpha}) = 0,
\end{equation}
where the $p$'s are polynomials in $\alpha$ (cf. \cite{PETKOVSEK1992243}). This finding was established by yet another application of the Mathematica FindSequenceFunction command. 

That is, we generated--for each value of $k$ under consideration--a sequence ($\alpha =1,2,\ldots,85$) of the rational values yielded by the hypergeometric-based formulas for $G_2^k(\alpha)$, to which the command was then applied.
While we have limited ourselves in App.~\ref{DiffEqs}
 to displaying our results 
for $k=-1,0,1, 2, 3$ and 4, we do have the analogous set of results in terms of the hypergeometric functions for 
the additional instances, $k=5,6,7,8$ and 9, and presume that an equivalent set of
difference-equation results is constructible (though substantial efforts with $k=5$ have not to this point succeeded).  The initial points $G_2^k(1)$ in the six difference equations shown are--in the indicated order--$\left\{\frac{1}{14},\frac{4}{33},\frac{45}{286},\frac{1553}{8398},\frac{3073}{14858},\frac{8348}{37145}\right\}$. The next five members of this monotonically-increasing sequence are 
$\left\{\frac{188373}{785726},\frac{1096583}{4342170},\frac{6050627}{22951470},\frac{160298199}{586426690},\frac{13988600951}{49611697974}\right\}$. 
Since, as noted above, $G_1^k(1)=1$, these are the respective separability probabilities $Q(k,1)$ themselves. We would like to extend this sequence sufficiently, so that we might be able to establish an underlying rule for it. (However, since the sequence is increasing in value, the Legendre-polynomial density-approximation procedure of Provost converges more slowly as $\alpha$ increases, so our quest seems somewhat problematical, despite the large number [15,801] of moments incorporated  [cf. \cite[App. II]{LatestCollaboration}].)

If in the difference equation for $k=-1$, we replace 
the term $G_2^{-1}(1)=\frac{1}{14}$ by $G_2^{-1}(1)=0$, then we can add
\begin{equation}
\frac{\pi  3^{-3 \alpha-5} 4^{3 \alpha+2} 5^{5 \alpha+3} \left(\frac{9}{10}\right)_{\alpha+1}
   \left(\frac{11}{10}\right)_{\alpha+1} \left(\frac{13}{10}\right)_{\alpha+1}
   \left(\frac{3}{2}\right)_{\alpha+1} \left(\frac{17}{10}\right)_{\alpha+1} \Gamma (\alpha) \Gamma
   (\alpha+2)}{52055003 \Gamma (5 \alpha) \Gamma \left(\alpha+\frac{1}{6}\right) \Gamma
   \left(\alpha+\frac{5}{6}\right)},
\end{equation}
to the $\alpha$-specific values obtained from the so-modified equation to recover the values generated by the original $k=-1$ difference equation.
\subsection{Polynomial coefficients in difference equations}
\subsubsection{The polynomials $p_2^k(\alpha)$}
We have for the six ($k=-1,0,1,2,3,4$) cases at hand  (App.~\ref{DiffEqs}) the proportionality
relation
\begin{equation} \label{p2Proportion}
p_2^k(\alpha) \propto \Pi_{i=1}^6 (u_{ik}-1),
\end{equation}
where the $u_{ik}$'s (and $b_{ik}$'s)--as indicated in sec.~\ref{Common}--are themselves functions of  $\alpha$.
\subsubsection{The polynomials $p_1^k(\alpha)$}
For all six displayed cases,
\begin{equation}
p_1^k(\alpha) \propto \Pi_{i=1}^6 b_{ik}  .
\end{equation}
\subsubsection{The polynomials $p_0^k(\alpha)$}
Further, for all six cases, the polynomial coefficients $p_0^k(\alpha)$--constituting the {\it inhomogeneous} parts of the recurrences--are proportional to the product of a factor of the form
\begin{equation}
\Pi_{i=1}^6 b_{ik} (b_{ik}-1),
\end{equation}
and an irreducible polynomial. These irreducible polynomials are, in the indicated order ($k=-1,0,1$),
\begin{equation} \label{p0minus1}
9250 \alpha ^4+12625 \alpha ^3+5645 \alpha ^2+938 \alpha +54,
\end{equation}
\begin{equation} \label{firstorderHS}
185000 \alpha ^5+779750 \alpha ^4+1289125 \alpha ^3+1042015 \alpha ^2+410694 \alpha
   +63000,
\end{equation}
\begin{equation} \label{firstorderk1}
74000 \alpha ^6+578300 \alpha ^5+1830820 \alpha ^4+3013197 \alpha ^3+2724024 \alpha
   ^2+1284280 \alpha +246960,
\end{equation}
and (for $k=2$)
\begin{equation}
740000 \alpha ^7+9002000 \alpha ^6+45576950 \alpha ^5+125164535 \alpha ^4 +202090226
   \alpha ^3
\end{equation}
\begin{displaymath}
+192332891 \alpha ^2+100092606 \alpha +22004136.
\end{displaymath}
 The irreducible polynomial for $k=3$ is also of degree 7, that is,
\begin{equation}
740000 \alpha ^7+11666000 \alpha ^6+76382750 \alpha ^5+271168745 \alpha ^4+566336789
   \alpha ^3\end{equation}
\begin{displaymath}
+698007782 \alpha ^2+471120306 \alpha +134548128.
\end{displaymath}
For $k=4$, this auxiliary polynomial $p^k_0(\alpha)$ is now the product of $(9 +4 \alpha)$ times an irreducible  polynomial of degree 7, that is,
\begin{equation}
296000 \alpha ^7+5584000 \alpha ^6+43492140 \alpha ^5+182972656 \alpha ^4+451645197
   \alpha ^3
\end{equation}
\begin{displaymath}
+656629192 \alpha ^2+522054355 \alpha +175452420.
\end{displaymath}
The coefficients of the highest powers of $\alpha$ in all six irreducible polynomials are factorable into the product of 37 and powers of 2 and 5.
\section{Hypergeometric-Free Formulas for $Q(k+1,\alpha)-Q(k,\alpha)$}
In App.~\ref{TermByTermDifferences} we show formulas we have 
generated for the differences between the formulas for $Q(k,\alpha)$ for successive
values of $k$. We note that these are hypergeometric-free.
We will find below (\ref{Q(k+1,a)Formula2}) that these obey the formula
\begin{equation} \label{Q(k+1,a)Formula}
Q(k+1,\alpha)-Q(k,\alpha)=
\end{equation}
\begin{displaymath}
\frac{\sqrt{\pi } 3^{-3 \alpha -1} \alpha  \Gamma \left(3 \alpha +\frac{3}{2}\right) (20
   \alpha +8 k+11) \Gamma \left(k+2 \alpha +\frac{3}{2}\right) \Gamma \left(k+3 \alpha
   +\frac{3}{2}\right) \Gamma (2 k+5 \alpha +2)}{2 \Gamma \left(\alpha
   +\frac{1}{2}\right) \Gamma \left(\alpha +\frac{5}{6}\right) \Gamma \left(\alpha
   +\frac{7}{6}\right) \Gamma (k+\alpha +2) \Gamma (k+4 \alpha +2) \Gamma \left(2 k+5
   \alpha +\frac{7}{2}\right)}.
\end{displaymath}
\section{Partial separability probability asymptotics}
\subsection{$k$-specific $\mbox{prob}(|\rho^{PT}| >|\rho|)$ formulas}
Now, as concerns the eleven formulas $Q(k,\alpha)$ ($k =-1,0,1,\ldots,9$) we have 
obtained for $\mbox{prob}(|\rho^{PT}| >|\rho|)$, which have been the principal focus of the paper, we have computed the ratios of the probability for  $\alpha=101$ to the 
probability for $\alpha =100$. These ranged from 0.419810 ($k=-1$) to 0.4204296 ($k=9$).
Let us note here that $z=\frac{27}{64} \approx 0.421875$.
\subsection{$\alpha$-specific $\mbox{prob}(|\rho^{PT}| >|\rho|)$ formulas} \label{kAsymptotics1}
We had available $\alpha =\frac{1}{2}, 1$ and 2 computations for $k=1,\ldots,40$ for this scenario. 
We found that, for each of the three values of $\alpha$, we could construct strongly linear plots--with unit-like slopes between 1.00177 and 1.00297--by
taking $k$ times the ratio ($R$) of the $(k+1)$ separability probability to the $k$-th separability probability. (From this, it appears, simply, that $R \rightarrow 1$, as 
$k \rightarrow \infty$.)
\subsection{``Diagonal'' $\alpha=k$ $\mbox{prob}(|\rho^{PT}| >|\rho|)$ formulas} \label{kAsymptotics2}
For values $\alpha =k =1,\ldots,50$, we were able to construct
a strongly linear plot by--similarly to the immediate last analysis--taking $k=\alpha$ times the ratio of the $(k+1)=(\alpha+1)$ separability probability to the $k=\alpha$-th separability probability. Now, however, rather than a slope very close to 1, we found a slope near to one-half, that is 0.486882. The ($k=\alpha =0$)-intercept of the estimated line was 0.894491.
\section{Total separability probability formulas}
Efforts of our to conduct parallel sets of ($k$-specific) analyses to those reported above
for the {\it total} separability probabilities $P(k,\alpha)$, corresponding to 
$|\rho^{PT}| > 0$, rather than for that component part
$Q(k,\alpha)$ of the probabilities satisfying the determinantal inequality 
$|\rho^{PT}| >
|\rho|$ had  been unsuccessful, in the following sense. We had computed what appeared to be appropriate sequences 
$(\alpha =1, 2,\ldots,74)$ of rational values for $k=1$ 
and $(\alpha =1, 2,\ldots,124)$ for $k=2$, 
but the Mathematica  FindSequenceFunction 
did not yield any underlying governing rules. (This can be contrasted with the results in \cite{LatestCollaboration}, where such  successes were reported in obtaining 
$\alpha$-specific [$|\rho^{PT}| > 0$] formulas [$\alpha = 1,2,\ldots,13$ and $\frac{1}{2}, \frac{3}{2},\frac{5}{2},\frac{7}{2}$], including (\ref{ComplexRule})-(\ref{RebitRule}) above. However, we do eventually succeed in characterizing the nature of these two  ($k=1,2$)  sequences [cf. sec.~\ref{AppTermByTermDifferencesFull}].) 

\begin{figure}
\includegraphics{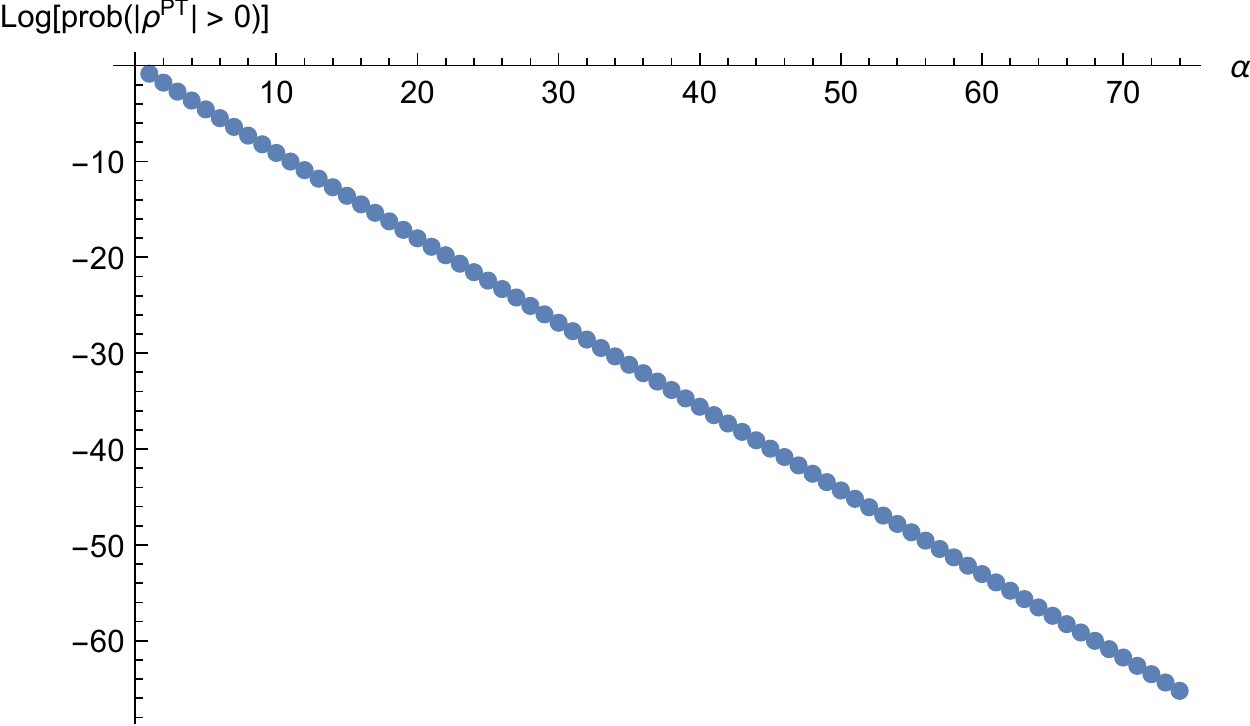}
\caption{\label{fig:Full}Plot of logs of total separability probability
($|\rho^{PT}| > 0$) for random induced measure with $k=1$. A least-squares linear fit to these 74 points is $-0.878482 \alpha -0.362781$.}
\end{figure}

\begin{figure}
\includegraphics{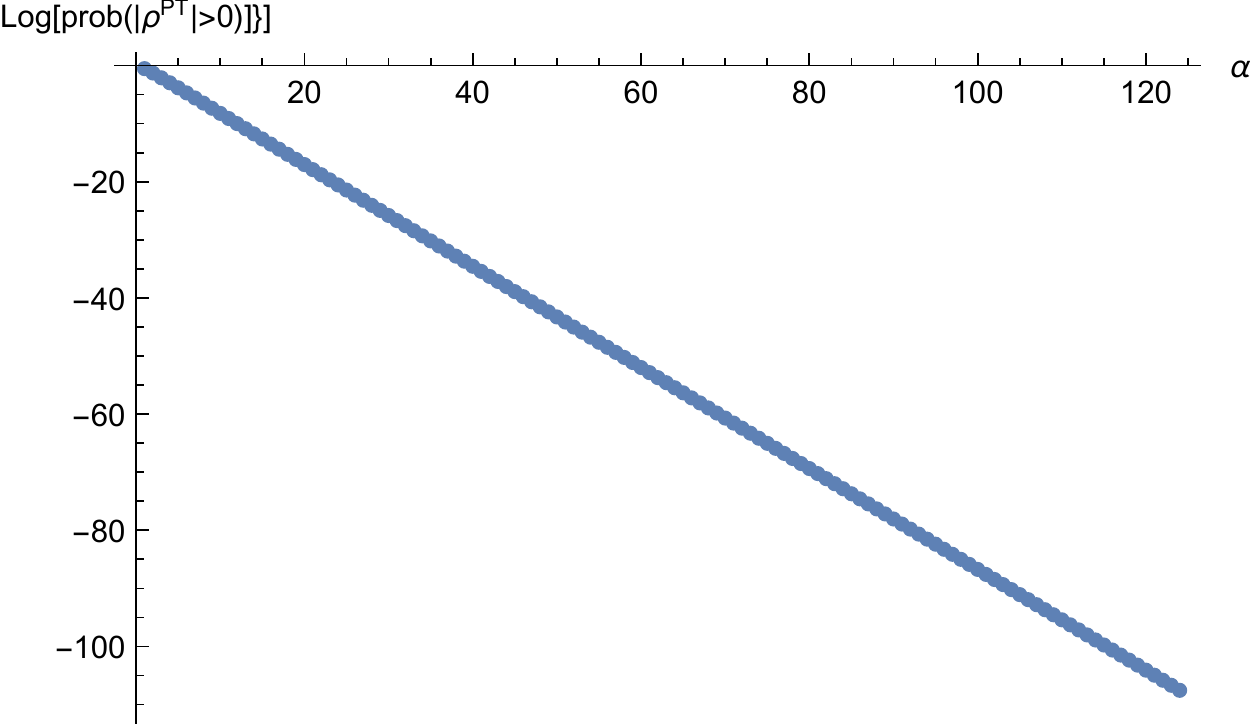}
\caption{\label{fig:Full2}Plot of logs of total separability probability
($|\rho^{PT}| > 0$) for random induced measure with $k=2$.  A least-squares linear fit to these 124 points is $-0.871033 \alpha + 0.351201$. We note that $\log \left(\frac{27}{64}\right) \approx -0.863046$.}
\end{figure}

In Fig.~\ref{fig:Full}, we plot the logs
of these $k=1$ seventy-four total separability probabilities (based on $\alpha=1,\ldots,74$). A least-squares linear fit to these points is $-0.878482 \alpha -0.362781$, while in Fig.~\ref{fig:Full2}, we show (based on $\alpha=1,\ldots,124$) the $k=2$ counterpart, with an analogous fit of $-0.871033 \alpha + 0.351201$. (We note that $\log \left(\frac{27}{64}\right) \approx -0.863046$.)
Although the slopes of these two linear fits are quite close, the $y$-intercepts themselves are of different sign. The predicted probabilities at $\alpha=1$, the first of the fitted points, are 0.289019 and 0.602955, respectively. In statistical parlance, the ``coefficients of determination'' or $R^2$ for the two linear fits to the log-plots are both greater than 0.99995. Further, sampling at $\alpha =1, 51, 101,\ldots, 1451$, we obtained an estimated, again, very-well fitting line of $-1.4754 -0.86417 \alpha$.
\subsection{Total separability probability asymptotics}
\subsubsection{$k$-specific $\mbox{prob}(|\rho^{PT}| >0)$ formulas}
C. Dunkl, on the basis of our $k =1, \alpha =1, 51, 101,\ldots, 1451$ analysis just above (and its companions), did
advance the bold and (certainly, in our overall analytical context) elegant  hypothesis of a $k$-invariant ($\alpha \rightarrow \infty$) slope equal to $\log{\frac{27}{64}} 
\approx -0.8630462173553$, which does seem quite consistent with the numerical properties
we have observed (that is, with the direction in which the estimates of the slope tend as the number of points sampled increase). 

As further support, we obtained for a  $k =2, \alpha =1, 49, 73 ,\ldots, 1465$ analysis, a slope estimate of -0.864025, again converging in the direction of 
$\log{\frac{27}{64}} $.
(Let us remark,  regarding the generalized two-qubit version of the [simpler, lower-dimensional] X-states model \cite{Xstates2,LatestCollaboration2,Beran}, that it has been  shown that the slope of a [now, log-log] plot 
of $\log({\mbox{prob}(|\rho^{PT}|>0})$ vs. $\log{\alpha}$ tends to $-\frac{1}{2}$, as $\alpha \rightarrow \infty$.)
\subsubsection{$\alpha$-specific $\mbox{prob}(|\rho^{PT}| >0)$ formulas} \label{kAsymptotics3}
These interesting observations led us to reexamine, for their asymptotic properties, the ``dual'' $P(k,\alpha)$ formulas (\ref{ComplexRule})-(\ref{RebitRule}), given above, and previously reported in \cite{LatestCollaboration}.
We now find--through analytic means--that for each of $\alpha = 1,2,3,4$ and $\frac{1}{2}, \frac{3}{2},\frac{5}{2},\frac{9}{2}$, that as $k \rightarrow \infty$, the ratio of the logarithm of the 
$(k+1)$-st separability probability to the logarithm of the 
$k$-th separability probability is $\frac{16}{27}$ (cf. \cite[sec. 7]{chu2014accelerating}). (Presumably, the pattern continues for larger $\alpha$, but the required computations have, so far, proved too challenging.)

\begin{figure}
\includegraphics{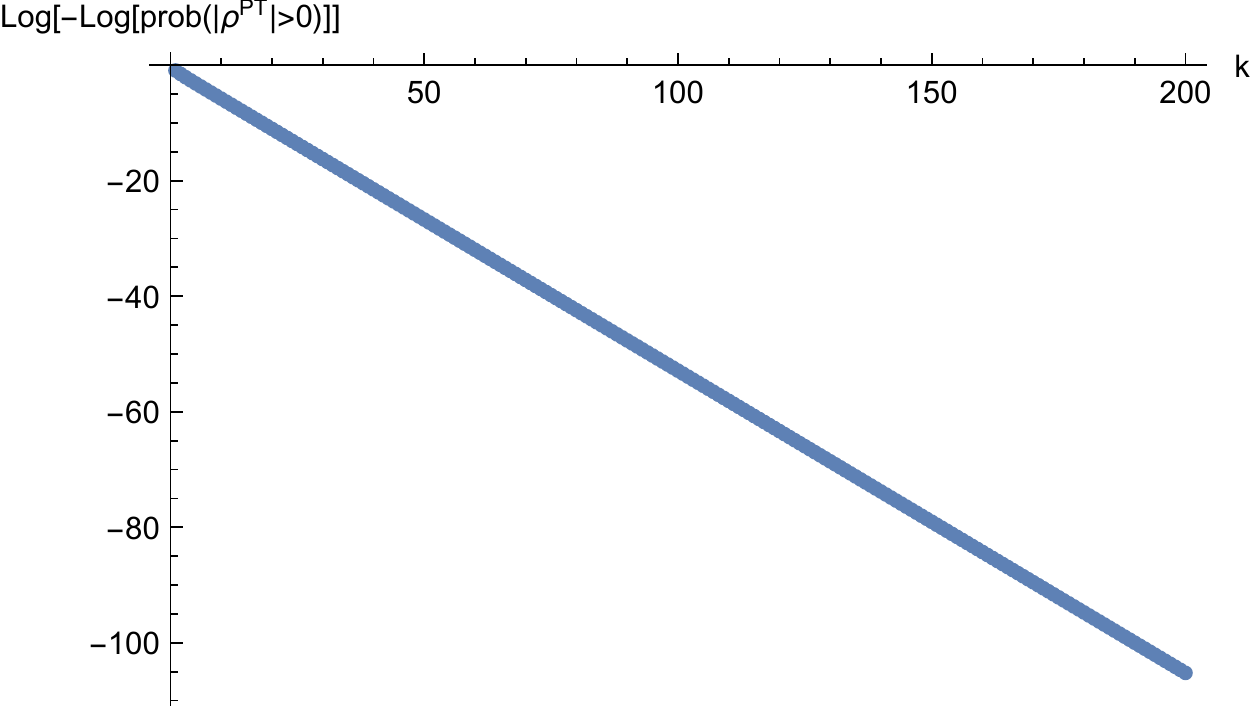}
\caption{\label{fig:Dual}Plot of $\log({-(\log{P(k,\frac{1}{2})}})$  vs. $k$. The slope of a least-squares-fitted line is 
-0.523280, while $\log{\frac{16}{27}} \approx -0.523248$.}
\end{figure}

For example, for $\alpha =\frac{1}{2}$, we have for the two-rebit total separability probability $P(k,\frac{1}{2})$, as a function of $k$, the formula (\ref{RebitRule}) given above.
In Fig.~\ref{fig:Dual}, we show a plot
of $\log({-(\log{P(k,\frac{1}{2})))}}$  vs. $k$. The slope of a 
least-squares-fitted line based on the 200 points is 
-0.523280, while $\log{\frac{16}{27}} \approx -0.523248$.
(As we increase $\alpha$ from $\frac{1}{2}$, but hold the number of points constant at 200, the approximation of the slope to this value slowly weakens.)
\section{``Concise formulas'' for $Q(k,\alpha)$}
Let us remind the reader of the interesting ``concise'' (Hilbert-Schmidt [$k=0$]) generalized two-qubit result--applying Zeilberger's (``creative telescoping'') algorithm \cite{doron}--of Qing-Hu Hou, reported in \cite[eqs. (1)-(3)]{slaterJModPhys}.
This--in our present notation--takes the form 
\begin{equation} \label{Hou1}
Q(0,\alpha) =\Sigma_{i=0}^\infty f_0(\alpha+i),
\end{equation}
where
\begin{equation} \label{Hou2}
f_0(\alpha) = Q(0,\alpha)-Q(0,\alpha +1) = \frac{ q_0(\alpha) 2^{-4 \alpha -6} \Gamma{(3 \alpha +\frac{5}{2})} \Gamma{(5 \alpha +2})}{6 \Gamma{(\alpha +1)} \Gamma{(2 \alpha +3)} 
\Gamma{(5 \alpha +\frac{13}{2})}},
\end{equation}
and
\begin{equation} \label{Hou3}
q_0(\alpha) = 185000 \alpha ^5+779750 \alpha ^4+1289125 \alpha ^3+1042015 \alpha ^2+410694 \alpha +63000 = 
\end{equation}
\begin{displaymath}
\alpha  \bigg(5 \alpha  \Big(25 \alpha  \big(2 \alpha  (740 \alpha
   +3119)+10313\big)+208403\Big)+410694\bigg)+63000.
\end{displaymath}
We divide the originally reported formula by one-half \cite{WholeHalf}, since we have moved here from the ($k=0$) Hilbert-Schmidt $|\rho^{PT}| >0$ original 
scenario to its  $|\rho^{PT}| > |\rho|$ counterpart.
Using our earlier results above, Hou has further been able to construct the $k=1$ analogue of the
``concise formula'' above (a Maple worksheet of his is presented in 
App.~\ref{AppendixHou} [ cf. \cite[Figs. 5, 6]{slaterJModPhys}). That is,
\begin{equation} \label{k1Hou1}
Q(1,\alpha) =\Sigma_{i=0}^\infty f_1(\alpha+i),
\end{equation}
where
\begin{equation}
f_1(\alpha)=  \frac{q_1(\alpha) \left(27 \right)^{\alpha } \Gamma (5 \alpha ) \Gamma \left(\alpha
   +\frac{5}{6}\right) \Gamma \left(\alpha +\frac{7}{6}\right)}{\left(50000 \right)^{\alpha }\Gamma (\alpha ) \Gamma
   \left(\alpha +\frac{17}{10}\right) \Gamma \left(\alpha +\frac{19}{10}\right) \Gamma
   \left(\alpha +\frac{21}{10}\right) \Gamma \left(\alpha +\frac{23}{10}\right) \Gamma
   (2 \alpha +5)}
\end{equation}
and
\begin{equation}
q_1(\alpha)=\frac{9 \pi}{1000000} (5 \alpha +1) (5 \alpha +2) (5 \alpha +3) \times
\end{equation}
\begin{displaymath}
 \left(74000 \alpha ^6+578300 \alpha ^5+1830820
   \alpha ^4+3013197 \alpha ^3+2724024 \alpha ^2+1284280 \alpha +246960\right).
\end{displaymath}
(These results correspond to the variable ``dif'' in App.~\ref{AppendixHou}.) Thus, in passing from the  (symmetric $k=0$) Hilbert-Schmidt setting to the random induced $k=1$ scenario, the degree of ``conciseness'' somewhat diminishes. 
The polynomials $q_0(\alpha)$ and $q_1(\alpha)$ in this pair of formulas are the same as the 
difference-equation (\ref{recurrence}) polynomials $p_0^0(\alpha)$ and 
$p_0^1(\alpha)$, given in (\ref{firstorderHS}) and (\ref{firstorderk1}).

At this point in our research, we were able to employ the Mathematica-based 
HolonomicFunctions package of Christoph Koutschan of the Research Institute for Symbolic Computation (RISC) of Johannes Kepler University. With it, we were readily able to derive
the $k=-1$ result
\begin{equation} \label{k1Houminus1}
Q(-1,\alpha) =P(-1,\alpha) =\Sigma_{i=0}^\infty f_{-1}(\alpha+i),
\end{equation}
where
\begin{equation}
f_{-1}(\alpha)= 
\end{equation}
\begin{displaymath}
 \frac{\pi  5^{-5 \alpha -4} 16^{-\alpha -1} 27^{\alpha } \left(\alpha  (10 \alpha +7)
   \left(925 \alpha ^2+615 \alpha +134\right)+54\right) \Gamma \left(\alpha
   +\frac{1}{6}\right) \Gamma \left(\alpha +\frac{5}{6}\right) \Gamma (5 \alpha
   +1)}{\Gamma \left(\alpha +\frac{9}{10}\right) \Gamma (\alpha +1) \Gamma \left(\alpha
   +\frac{11}{10}\right) \Gamma \left(\alpha +\frac{13}{10}\right) \Gamma \left(\alpha
   +\frac{17}{10}\right) \Gamma (2 \alpha +2)}.
\end{displaymath}
We see
that the polynomial $\alpha  (10 \alpha +7) \left(925 \alpha ^2+615 \alpha +134\right)+54$ above is, in expanded form, the same as $p_0^{-1}(\alpha)$ given in (\ref{p0minus1}).

For the standard trio of Dyson-indices $\alpha = \frac{1}{2}, 1$ and 2, this formula  for $Q(-1,\alpha)$ yields $\frac{1}{8}, \frac{1}{14}$ and $\frac{11}{442}$, respectively, while $\alpha=-\frac{1}{2}, 0$ lead to $\frac{1}{3}, \frac{1}{5}$. (Also, $\alpha =-\frac{3}{2}$ gives $\frac{1}{3}$, and $\alpha=-1$ yields $\frac{1}{5}$.) Additionally, $\alpha=-\frac{1}{3}$ gives $\frac{19}{60} C^2$, where $C^2$ is the Baxter's four-coloring constant for a triangular lattice, that is, $C^2=\frac{3}{4 \pi^2} \Gamma \left(\frac{1}{3}\right)^3$. (Also, $\alpha =\frac{2}{3}$ gives $1-\frac{27 C^2}{44}$.) Continuing with this ``zoo'' of remarkable results (suggested largely by use of WolframAlpha), $\alpha=\frac{1}{4}$
gives $1-G_{GA} \approx 0.1653731583$, where $G_{GA}$ is Gauss's constant, that is, the 
reciprocal of the {\it arithmetic-geometric mean} of 1 and $\sqrt{2}$, equalling 
$\frac{\Gamma \left(\frac{1}{4}\right)^2}{2 \sqrt{2} \pi ^{3/2}}$. Now, for $\alpha=-\frac{1}{4}$, we get $\frac{8}{5 L}+1 \approx 1.6102078108$, where $L$ is the Lemniscate constant, that is, $L=\frac{1}{2 \sqrt{2 \pi}} \Gamma \left(\frac{1}{4}\right)^2$. To continue, $\alpha = -\frac{2}{3}$ gives us  
$1-\frac{163}{1008 \Im\left(\omega _1\right)} \approx 0.87795554$, where $\omega _1 = \frac{(1+i \sqrt{3}) 
\Gamma \left(\frac{1}{3}\right)^3}{8 \pi}$, is a known constant of interest 
(cf. \cite[sec. 3.2.1]{slaterJModPhys}).

Further, employing the RISC package, we obtained
\begin{equation} \label{Koutschank3}
Q(3,\alpha)  =\Sigma_{i=0}^\infty f_{3}(\alpha+i),
\end{equation}
where
\begin{equation}
f_{3}(\alpha)= 
\end{equation}
\begin{displaymath}
\frac{3^{3 \alpha +4} 4^{-2 \alpha -5} (2 \alpha +5) \Gamma \left(\alpha
   +\frac{8}{5}\right) \Gamma \left(\alpha +\frac{9}{5}\right) \Gamma \left(\alpha
   +\frac{11}{6}\right) \Gamma \left(\alpha +\frac{13}{6}\right) \Gamma \left(\alpha
   +\frac{11}{5}\right) \Gamma \left(\alpha +\frac{12}{5}\right) q_3(\alpha )}{625
   \sqrt{5} \pi  (\alpha +4) \Gamma \left(\alpha +\frac{27}{10}\right) \Gamma
   \left(\alpha +\frac{29}{10}\right) \Gamma \left(\alpha +\frac{31}{10}\right) \Gamma
   \left(\alpha +\frac{33}{10}\right) \Gamma (2 \alpha +7)},
\end{displaymath}
and
\begin{displaymath}
q_3(\alpha)=
\end{displaymath}
\begin{displaymath}
\alpha  (\alpha  (\alpha  (5 \alpha  (50 \alpha  (8 \alpha  (370 \alpha
   +5833)+305531)+54233749)+566336789)+698007782)+471120306)+134548128
\end{displaymath}
is a degree-7 polynomial in $\alpha$.

For the standard trio of Dyson-indices $\alpha = \frac{1}{2}, 1$ and 2, this formula  for $Q(3,\alpha)$ yields $\frac{84883}{262144},  \frac{3073}{14858}$ and $\frac{3439}{41354}$, respectively.

It would clearly be of interest to find such ``concise'' expressions for $Q(k,\alpha)$, encompassing the
four ($k=-1,0,1,3$) examples above, as well as values $k>3$.  (We have so far 
encountered certain difficulties in applying the RISC HolonomicFunctions program to the $k=2$ scenario.)
\section{Series of exact $k$-values for certain $\alpha$ and associated formulas}
\subsection{Series}
We have previously noted $Q(-1,-\frac{1}{3})=\frac{19}{60} C^2$, where $C^2$ is the Baxter's four-coloring constant for a triangular lattice, that is, $C^2=\frac{3}{4 \pi^2} \Gamma \left(\frac{1}{3}\right)^3$. For the succeeding values $k=0,\dots,9$, we obtain
$\frac{C^2}{2}+1,1-\frac{3 C^2}{20},1-\frac{783 C^2}{3740},1-\frac{1171341
   C^2}{4989160},1-\frac{51068151 C^2}{204555560},1-\frac{132326834139
   C^2}{509547899960},\newline 1-\frac{8028455705181 C^2}{30063326097640}, 1-\frac{582160729281381
   C^2}{2134496152932440},1-\frac{4372426421400790827
   C^2}{15767523081711934280},1-\frac{447620586926496661827
   C^2}{1592519831252905362280}$.

For the series ($k=-1,0,\ldots9$) with $\alpha=-\frac{1}{2}$, we obtain $\left\{\frac{1}{3},\frac{1}{3},1,\frac{13}{16},\frac{191}{256},\frac{1453}{2048},\frac{4
   4923}{65536},\frac{350323}{524288},\frac{5494379}{8388608},\frac{43249277}{67108864},
   \frac{2730885203}{4294967296}\right\}$. Here, all the denominators ($k=1,\ldots,9$) are simply increasing powers of 2.

For the series ($k=0,\ldots9$) with $\alpha=-\frac{1}{4}$, we obtain 
$\let\left\relax\let\right\relax
\left\{1,1-\frac{4}{5 L},1-\frac{184}{195 L},1-\frac{1116}{1105 L},1-\frac{504688}{480675
   L},1-\frac{19161148}{17784975 L},1-\frac{47082376}{42893175
   L},1-\frac{301219589404}{270527254725 L},1-\frac{18584575275424}{16502162538225
   L},1-\frac{288596382356}{253879423665 L}\right\}$,
where $L$ is the indicated Lemniscate constant, that is, $L=\frac{1}{2 \sqrt{2 \pi}} \Gamma \left(\frac{1}{4}\right)^2$.
\subsection{Formulas}
This last series has the explanatory rule ($k=0,1,\ldots9$)
\begin{equation} \label{Q-1/4}
Q(k,-\frac{1}{4})= \frac{\Gamma \left(\frac{5}{4}\right) \left(\frac{\Gamma \left(2 k+\frac{3}{4}\right) \,
   _3F_2\left(1,k+\frac{3}{8},k+\frac{7}{8};k+\frac{9}{8},k+\frac{13}{8};1\right)}{\Gamma
   \left(2 k+\frac{9}{4}\right)}-\sqrt{\pi }\right)}{L \Gamma \left(\frac{3}{4}\right)}+1=
\end{equation}
\begin{displaymath}
2^{-2 k-\frac{9}{4}} \Gamma \left(2 k+\frac{3}{4}\right) \,
   _3\tilde{F}_2\left(1,k+\frac{3}{8},k+\frac{7}{8};k+\frac{9}{8},k+\frac{13}{8};1\right)
   +\frac{1}{2},
\end{displaymath}
where the {\it regularized} hypergeometric function is indicated.
For $k=-1$, the formula  yields $1+\frac{4}{5 L}$, while our prior computations
indicate a value of $1+\frac{8}{5 L}$.

Also (now agreeing for $k=-1,0,\ldots,9$),
\begin{equation} 
Q(k,\frac{1}{4})= 1+ \frac{L}{21 \pi} U
\end{equation}
where
\begin{displaymath}
U=4 \, _3F_2\left(\frac{5}{8},1,\frac{9}{8};\frac{11}{8},\frac{15}{8};1\right)-21-
\end{displaymath}
\begin{displaymath}
\frac{4 \Gamma \left(\frac{11}{4}\right) \Gamma \left(2 k+\frac{13}{4}\right) \,
   _3F_2\left(1,k+\frac{13}{8},k+\frac{17}{8};k+\frac{19}{8},
k+\frac{23}{8};1\right)}{\Gamma \left(\frac{5}{4}\right) \Gamma \left(2 k+\frac{19}{4}\right)}.
\end{displaymath}
Absorbing the Lemniscate constant $L$, we obtain, equivalently, 
\begin{displaymath}
Q(k,\frac{1}{4})= -2^{-2 k-\frac{19}{4}} \Gamma \left(2 k+\frac{13}{4}\right) \,
   _3\tilde{F}_2\left(1,k+\frac{13}{8},k+\frac{17}{8};k+\frac{19}{8},k+\frac{23}{8};1
\right)+ \frac{1}{2}.
\end{displaymath}
We see some obvious parallels between the formulas for $Q(k,-\frac{1}{4})$ and  $Q(k,\frac{1}{4})$. (We note that $Q(0,\frac{1}{4})-Q(k,-\frac{1}{4}) =-\frac{17 G_{Ga}}{21}$, where Gauss's constant is indicated.)

In fact, we can subsume both these last two formulas ($\alpha =-\frac{1}{4}, \frac{1}{4}$) into
\begin{equation} \label{generalized}
Q(k,\alpha)=
\end{equation}
\begin{displaymath}
\frac{1}{2}-2^{-5 \alpha -2 k-\frac{7}{2}} \text{sgn}(\alpha ) \Gamma (2 k+5 \alpha +2)
   \, _3\tilde{F}_2\left(1,k+\frac{5 \alpha }{2}+1,k+\frac{5 \alpha
   }{2}+\frac{3}{2};k+\frac{5 \alpha }{2}+\frac{7}{4},k+\frac{5 \alpha
   }{2}+\frac{9}{4};1\right).
\end{displaymath}

Building upon (\ref{generalized}), we found
\begin{equation}
Q(k,\frac{3}{4})=
\end{equation}
\begin{displaymath}
-2^{-2 k-\frac{29}{4}} \Gamma \left(2 k+\frac{23}{4}\right) \,
   _3\tilde{F}_2\left(1,k+\frac{23}{8},k+\frac{27}{8};k+\frac{29}{8},k+\frac{33}{8};1\right)-\frac{\Gamma \left(2 k+\frac{23}{4}\right)}{2 \sqrt{\pi } (k+3) \Gamma \left(2
   k+\frac{21}{4}\right)}+\frac{1}{2}.
\end{displaymath}

Strikingly simply, we have the result (valid for all eleven values $k=-1,0,\ldots,9$ for which we have computations)
\begin{equation} \label{Q1/2}
Q(k,\frac{1}{2})=\frac{1}{2}-\frac{\Gamma \left(2 k+\frac{9}{2}\right)}{\sqrt{\pi } \Gamma (2 k+5)}
\end{equation}
(having a root at $k=-\frac{3}{2}$).
So, using formula (\ref{RebitRule}) above, we find that the {\it complementary} separability probability, that is, that associated with the determinantal inequality
$|\rho| > |\rho^{PT}| \geq  0$ is 
\begin{equation}
P(k,\frac{1}{2})-Q(k,\frac{1}{2})= \frac{\Gamma \left(2 k+\frac{9}{2}\right) \left(\frac{1}{\Gamma (2 k+5)}-\frac{4^{k+1} (8
   k+15) \Gamma (k+2)}{\Gamma (3 k+7)}\right)}{\sqrt{\pi }}+\frac{1}{2}.
\end{equation}
Also, we have found (agreeing with the earlier formulas for all eleven $k$) that 
\begin{equation}
Q(k,-\frac{1}{2}) = \frac{\Gamma \left(2 k-\frac{1}{2}\right)}{\sqrt{\pi } \Gamma (2 k)}+\frac{1}{2},
\end{equation}
for $k=1,2,\ldots9$, with the results for $k=-1,0$ of $\frac{1}{2}$ differing 
from the prediction of $\frac{1}{3}$ given by the early formulas given above.

Further, we have
\begin{equation}
Q(k,1)=\frac{1}{2}-\frac{4^{k+3} \Gamma \left(k+\frac{7}{2}\right)^2 \Gamma
   \left(k+\frac{9}{2}\right)}{\pi  \Gamma (k+5) \Gamma \left(2 k+\frac{13}{2}\right)},
\end{equation}
having a root at $k=-2$.

To continue (with a root at $k=-\frac{5}{2}$),
\begin{equation}
Q(k,\frac{3}{2})=\frac{1}{2}-\frac{(6 k+31) \Gamma \left(2 k+\frac{19}{2}\right)}{4 \sqrt{\pi } (k+5)
   (k+6) \Gamma (2 k+9)}.
\end{equation}

Further,
\begin{equation}
Q(k,2)= \frac{1}{2}-\frac{4^{k+6} (k+6) \Gamma \left(k+\frac{11}{2}\right) \Gamma
   \left(k+\frac{13}{2}\right) \Gamma \left(k+\frac{15}{2}\right)}{\pi  \Gamma (k+9)
   \Gamma \left(2 k+\frac{23}{2}\right)}
\end{equation}
(having a root at $k=-3$) agreeing with our earlier formulas for all eleven $k$ (as well as $k=-2$ and 10).

Our formulas give that $Q(k,0)$ is equal to $\frac{1}{5}$ for both $k=-2$ and 1, and equal to $\frac{1}{2}$ for $k=0,\ldots,9$. Here, $\alpha = 0 $ presumably corresponds to a classical/nonquantum scenario.

Charles Dunkl has observed that for integral values of $\alpha$, the arguments of the 
gamma functions in the numerators are of the form $\left\{2 \alpha +k+\frac{3}{2},2 \alpha +\left\lfloor \frac{\alpha }{2}\right\rfloor
   +k+\frac{3}{2},3 \alpha +k+\frac{3}{2}\right\}$, and in the denominators of the form
$\left\{k+4 n+1,2 k+5 n+\frac{3}{2}\right\}$. He further noted that the leading
(highest power) in the polynomial takes the form $\alpha  2^{5 \alpha +2 k+1} k^{\alpha +\left\lfloor \frac{\alpha -1}{2}\right\rfloor -1}$. Also, the second leading coefficient
(normalizing the leading coefficient of the polynomial to 1) follows the rule
\begin{equation}
c_2=\frac{1}{48} \left(190 \alpha ^2-174 \alpha -3 (-1)^{\alpha } (10 \alpha +3)-55\right).
\end{equation}
Similarly, the so-normalized leading third coefficient takes the form
\begin{equation}
c_3=
\end{equation}
\begin{displaymath}
\frac{30766 \alpha ^4-77260 \alpha ^3+23350 \alpha ^2-5 (-1)^{\alpha } \left(2 \alpha 
   \left(950 \alpha ^2-885 \alpha -716\right)-213\right)+26920 \alpha +3799}{3840}.
\end{displaymath}

We have been able to generate a considerable number (including $k =1,\ldots,100$) of such $Q(k,\alpha)$ formulas, a limited number of which
we present  in App.~\ref{AppendixFormulas}.

Each half-integral $\alpha$ formula contains a gamma function in its numerator with an argument of the 
form $2 +5 \alpha + 2 k$ and in its denominator a gamma function with an argument of the form $2 k+\frac{1}{2} (-1)^{\alpha } \left(-2 (-1)^{\alpha } (5 \alpha +2)-i\right)$.
\subsection{Sets of consecutive negative roots} \label{consecutiveroots}
All the $Q(k,\alpha)$ formulas we have (App.~\ref{AppendixFormulas}), for
nonnegative half-integer and integer values of $\alpha$, have roots (in unit steps) from $k=-\alpha-1$ downwards to $k=-\frac{1}{4} (-1)^\alpha \left((-1)^\alpha (10 \alpha+1)-1\right)$. So, there 
are 
\begin{equation} \label{NumberRoots}
-\alpha +\frac{1}{4} (-1)^{\alpha } \left((-1)^{\alpha } (10 \alpha +1)-1\right)-1
\end{equation} associated roots. (The formulas displayed in App.~\ref{AppendixFormulas} with negative values of $\alpha$ match our computations only above certain [nonnegative] values of $k$.)

\section{Hypergeometric formula for $Q(k,\alpha)$} \label{DunklInsight}
Based on the information presented above, including that in an extended form of
App.~\ref{AppendixFormulas}, C. Dunkl developed the following formula, succeeding in reproducing our computations for $\alpha=0,1,2,\ldots$
\begin{equation} \label{CharlesInteger}
Q(k,\alpha)= Q(-\alpha ,\alpha ) \sum _{j=0}^{\alpha +k} H(\alpha ,j)
\end{equation}
where
\begin{equation} \label{Qaa}
Q(-\alpha,\alpha)= \frac{1}{2} (\frac{4}{27})^\alpha \frac{\left(\frac{3}{4}\right)_{\alpha }
   \left(\frac{5}{4}\right)_{\alpha }}{\left(\frac{5}{6}\right)_{\alpha }
   \left(\frac{7}{6}\right)_{\alpha }}
\end{equation}
and 
\begin{displaymath}
H(\alpha,j)=\frac{\left(\frac{3 \alpha }{2}\right)_j \left(\alpha +\frac{1}{2}\right)_j \left(\frac{3
   \alpha }{2}+\frac{1}{2}\right)_j \left(\frac{3 \alpha }{2}+\frac{11}{8}\right)_j
   \left(2 \alpha +\frac{1}{2}\right)_j}{j! \left(\frac{3 \alpha
   }{2}+\frac{3}{8}\right)_j \left(\frac{3 \alpha }{2}+\frac{3}{4}\right)_j \left(\frac{3
   \alpha }{2}+\frac{5}{4}\right)_j (3 \alpha +1)_j}.
\end{displaymath}
(In explaining how this formula was obtained, Dunkl stated that the key insights
was that $Q(k+1,\alpha)-Q(k,\alpha)$ factors nicely and that $Q(-\alpha-1,\alpha)=0$.)
If we let both $\alpha$ and $k$ be free, and perform the indicated summation in 
(\ref{CharlesInteger}), we obtain a hypergeometric-based formula that appears not only to reproduce the formulas in App.~\ref{AppendixFormulas} for integer $\alpha$, but also
half-integer and other nonnegative fractional values (such as $\frac{1}{4}, \frac{2}{3}$) of $\alpha$.

Dunkl argued that for 
$k>-\alpha$ and $n=1,2,3,\ldots$%
\begin{align*}
Q\left(  k+n,\alpha\right)    & =Q\left(  k,\alpha\right)  +\left(  Q\left(
k+1,\alpha\right)  -Q\left(  k,\alpha\right)  \right)  +\left(  Q\left(  k+2,\alpha\right)
-Q\left(  k+1,\alpha\right)  \right)  +\cdots\\
& +\cdots+\left(  Q\left(  k+n,\alpha\right)  -Q\left(  k+n-1,\alpha\right)  \right)
\\
& =Q\left(  k,\alpha\right)  +Q\left(  -\alpha,\alpha\right)  \sum_{i=0}^{n-1}H\left(
\alpha,k+\alpha+1+i\right)  .
\end{align*}
Taking the limit as $n\rightarrow\infty$%
\begin{align*}
\frac{1}{2}  & =Q\left(  k,\alpha\right)  +Q\left(  -\alpha,\alpha\right)  \sum_{i=0}%
^{\infty}H\left(  \alpha,k+\alpha+1+i\right)  \\
& =Q\left(  k,\alpha\right)  +Q\left(  -\alpha,\alpha\right)  H\left(  \alpha,k+\alpha+1\right)
\sum_{i=0}^{\infty}\frac{H\left(  \alpha,k+\alpha+1+i\right)  }{H\left(  \alpha,k+\alpha+1\right)
},
\end{align*}
thus%
\[
Q\left(  k,\alpha\right)  =\frac{1}{2}-Q\left(  -\alpha,\alpha\right)  H\left(
\alpha,k+\alpha+1\right)  \sum_{i=0}^{\infty}\frac{H\left(  \alpha,k+\alpha+1+i\right)  \left(
1\right)  _{i}}{H\left(  \alpha,k+\alpha+1\right)  i!}.%
\]
(Let us point the reader to an interesting partial matching between entries of the hypergeometric function and arguments of the gamma functions.) 
The resultant master formula takes the form
\begin{align*} \label{Hyper1}
Q\left(  k,\alpha\right)   &  =\frac{1}{2}-\frac{\alpha\left(  20\alpha+8k+11\right)
\Gamma\left(  5\alpha+2k+2\right)  \Gamma\left(  3\alpha+k+\frac{3}{2}\right)
\Gamma\left(  2\alpha+k+\frac{3}{2}\right)  }{4\sqrt{\pi}\Gamma\left(
5\alpha+2k+\frac{7}{2}\right)  \Gamma\left(  \alpha+k+2\right)  \Gamma\left(
4\alpha+k+2\right)  }\\
&  \times~_{6}F_{5}\left(
\genfrac{}{}{0pt}{}{1,\frac{5}{2}\alpha+k+1,\frac{5}{2}\alpha+k+\frac{3}{2}%
,2\alpha+k+\frac{3}{2},3\alpha+k+\frac{3}{2},\frac{5}{2}\alpha+k+\frac{19}{8}%
}{\alpha+k+2,4\alpha+k+2,\frac{5}{2}\alpha+k+\frac{7}{4},\frac{5}{2}\alpha+k+\frac{9}{4},\frac
{5}{2}\alpha+k+\frac{11}{8}}%
;1\right).
\end{align*}
The value $\frac{1}{2}$ from which these terms are subtracted itself has an interesting 
provenance. It was obtained by conducting the sum indicated in (\ref{CharlesInteger}), not over $j$ from 0 to $\alpha+k$ as indicated there, but over $j$ from 0 to $\infty$, that is $ Q(-\alpha ,\alpha ) \sum _{j=0}^{\infty} H(\alpha ,j)$. 
(The $Q\left(  k,\alpha\right)$ formula can then be recovered by subtracting the sum over $j$ from $\alpha+k+1$ to $\infty$, that is, $ Q(-\alpha ,\alpha ) \sum _{j=\alpha +k+1}^{\infty} H(\alpha ,j)$.) This resulted in the expression (cf. http://math.stackexchange.com/questions/1872364/prove-that-a-certain-hypergeometric-function-assumes-either-the-value-frac1)
\begin{equation}\label{Hyper2}
Q(-\alpha ,\alpha ) \sum _{j=0}^{\infty} H(\alpha ,j)=
\end{equation}
\begin{displaymath}
\frac{\sqrt{\pi } 3^{-3 \alpha -1} \Gamma \left(2 \alpha +\frac{3}{2}\right) \,
   _5F_4\left(\frac{3 \alpha }{2},\alpha +\frac{1}{2},\frac{3 \alpha
   }{2}+\frac{1}{2},\frac{3 \alpha }{2}+\frac{11}{8},2 \alpha +\frac{1}{2};\frac{3 \alpha
   }{2}+\frac{3}{8},\frac{3 \alpha }{2}+\frac{3}{4},\frac{3 \alpha }{2}+\frac{5}{4},3
   \alpha +1;1\right)}{\Gamma \left(\alpha +\frac{5}{6}\right) \Gamma \left(\alpha
   +\frac{7}{6}\right)}.
\end{displaymath}
For $\alpha>0$ this gives us the indicated value of $\frac{1}{2}$. Let us note that for both this $_5F_4$ function and the $_6F_5$ immediately preceding, the sums of the denominator entries minus the sums of the numerator parameters equal $\frac{1}{2}$--while if these differences had been 1, the two functions could be designated as ``$\frac{1}{2}$-balanced'' \cite{wenchang2002}.

In the notation of this section (cf. (\ref{Q(k+1,a)Formula})),
\begin{equation} \label{Q(k+1,a)Formula2}
Q(k+1,\alpha)-Q(k,\alpha) = Q(-\alpha,\alpha) H(\alpha,\alpha+k+1)=
\end{equation}
\begin{displaymath}
\frac{\sqrt{\pi } 3^{-3 \alpha -1} \alpha  \Gamma \left(3 \alpha +\frac{3}{2}\right) (20
   \alpha +8 k+11) \Gamma \left(k+2 \alpha +\frac{3}{2}\right) \Gamma \left(k+3 \alpha
   +\frac{3}{2}\right) \Gamma (2 k+5 \alpha +2)}{2 \Gamma \left(\alpha
   +\frac{1}{2}\right) \Gamma \left(\alpha +\frac{5}{6}\right) \Gamma \left(\alpha
   +\frac{7}{6}\right) \Gamma (k+\alpha +2) \Gamma (k+4 \alpha +2) \Gamma \left(2 k+5
   \alpha +\frac{7}{2}\right)}.
\end{displaymath}
\subsection{Implications for $P(k,\alpha)$ formula}
Let us note that for the Hilbert-Schmidt ($k=0$) case, apparently \cite{WholeHalf}, $2 Q(0,\alpha) =P(0,\alpha)$, where 
\begin{equation}
Q(0,\alpha)=\frac{1}{2}-
\end{equation}
\begin{displaymath}
 \resizebox{1.10\hsize}{!}{%
        $\frac{2^{-4 \alpha -4} (20 \alpha +11) \Gamma \left(3 \alpha
   +\frac{3}{2}\right) \Gamma (5 \alpha +2) \, _6F_5\left(1,2 \alpha +\frac{3}{2},\frac{5
   \alpha }{2}+1,\frac{5 \alpha }{2}+\frac{3}{2},\frac{5 \alpha }{2}+\frac{19}{8},3
   \alpha +\frac{3}{2};\alpha +2,\frac{5 \alpha }{2}+\frac{11}{8},\frac{5 \alpha
   }{2}+\frac{7}{4},\frac{5 \alpha }{2}+\frac{9}{4},4 \alpha +2;1\right)}{\Gamma (2
   \alpha ) \Gamma (\alpha +2) \Gamma \left(5 \alpha +\frac{7}{2}\right)}$%
	}.
\end{displaymath}
Thus, any presumed ``master formula'' for $P(k,\alpha)$ (sec.~\ref{MasterP}), should reduce to $2 Q(0,\alpha)$ for $k=0$ (cf. eqs. 
(\ref{Hou1})-(\ref{Hou3})). We have been investigating the use of $2 Q(k,\alpha)$ as an initial candidate for $P(k,\alpha)$, then padding out the six upper and five lower entries of the 
$_6F_5$ function with additional pairs of entries, identical for $k=0$, but different for $k \neq 0$. Then, for $k=0$, the initial candidate would be recovered. (The somewhat interesting ``$\frac{1}{2}$-balanced'' property, mentioned above, or some $k$-free counterpart of it would, then, be lost.) Initial limited numerical investigations along these lines have been somewhat disappointing, as they appeared to indicate that the best fits
would be obtained for pairs of padded entries with {\it equal} coefficients of $k$. Also, fits to values of $P(k,\alpha)$ did not seem to be improved through the padding strategy.

However, another considerably more interesting approach along similarly motivated lines was, then, developed. We mapped the parameter $k$ in the $Q(k,\alpha)$ function to $\beta k$, so that for $k=0$ the original function would be recovered, no matter the specific value of $\beta$. We evaluated the transformed functions by seeing how well they fit the series of (known) eight values 
$P(k,k)$, $k =5,\ldots,12$. For the original 
$\beta= 1$, the figure-of-merit for the fit was 0.7703536. This figure rather dramatically decreases/improves as $\beta$ increases, reaching  a near minimum of 0.0479732 for 
$\beta =\frac{11}{2}$ (and 0.108008 for $\beta =5$ and 0.153828 for $\beta=6$.) The implications of this phenomenon will be further investigated.
Perhaps it might be of value to {\it combine} the last two (padding and scaling of $k$) strategies.
\subsection{Conjectured Identity}
In relation to (\ref{Hyper2}), Dunkl  formulated the conjecture 
\begin{equation} \label{Conjectured Hypergeometric Identity}
_5F_4\left(\frac{3 \alpha }{2},\alpha +\frac{1}{2},\frac{3 \alpha
   }{2}+\frac{1}{2},\frac{3 \alpha }{2}+\frac{11}{8},2 \alpha +\frac{1}{2};\frac{3 \alpha
   }{2}+\frac{3}{8},\frac{3 \alpha }{2}+\frac{3}{4},\frac{3 \alpha }{2}+\frac{5}{4},3
   \alpha +1;1\right) 
\end{equation}
\begin{displaymath}
= \frac{3}{2 \sqrt{2}} (\frac{27}{4})^{\alpha} \frac{\Gamma \left(\alpha +\frac{5}{6}\right) \Gamma \left(\alpha
   +\frac{7}{6}\right)}{\Gamma \left(\alpha +\frac{3}{4}\right) \Gamma
   \left(\alpha +\frac{5}{4}\right)}.
\end{displaymath}
To avoid zero denominators, it is necessary that $\alpha > -\frac{1}{8}$. For $\alpha=0$, the value is 1, while the sum is rational for $\alpha=n,n+\frac{1}{2}$, $n=0,1,2\ldots$.

In response to this conjecture, C. Koutschan wrote: ``The 5F4 sum fits into the class of identities
that can be done with Zeilberger's algorithm. I attach a Mathematica
notebook  with some computations. More precisely,
using the creative telescoping method, my program finds a linear
recurrence equation that is satisfied by the 5F4 sum. It is a trivial calculation
to verify that also the right-hand side satisfies the same recurrence.
As you remark, both sides give 1 for $\alpha=0$. We can conclude that the
identity holds for all $\alpha$  in $\mathbb{N}$.'' However, cases where $\alpha$ is neither an integer or half-integer still require attention. (G. Gasper has commented that the $_{5}F_{4}$ function is {\it not} a special case of the formulas in his paper
with M. Rahman \cite{GasperRahman}.)
\section{Master Formula Investigation for $P(k,\alpha)$} \label{MasterP}
Appendix A in \cite{LatestCollaboration} considered the possibility of developing a master formula for the total separability probability $P(k,\alpha)$, that associated with the determinantal inequality
$|\rho^{PT}|>0$ (cf. (\ref{ComplexRule})-(\ref{RebitRule})). 
It now clearly seems appropriate to reexamine those results (App.~\ref{AppendixFormulasTotalProbs}) in terms of the striking hypergeometric-based formula (sec.~\ref{DunklInsight}) we have obtained for the partial separability probability $Q(k,\alpha)$, that associated with the determinantal inequality
$|\rho^{PT}|>|\rho|$. 

In the earlier study \cite{LatestCollaboration}, the formulas took the form of 1 minus terms involving polynomials in $k$ and gamma functions, while above the interesting such terms have been subtracted from 
$\frac{1}{2}$. So, conjecturally there exists a tightly-related analogue of the results reported in 
sec.~\ref{DunklInsight} for $P(k,\alpha)$. (Dunkl did note the qualitative difference that
``the ratio $\frac{\frac{1}{2} - Q(k+1,\alpha)}{\frac{1}{2}-Q(k,a)}$ tends to 1 as $k \rightarrow \infty$ but $\frac{1 - P(k+1,\alpha)}{1-P(k,a)}$ tends to $\frac{16}{27}$.'') 

In investigating these matters, we have found that for our set of
computed $P(k,\alpha), \alpha =1,\ldots,47$, the number and location of the consecutive negative roots (sec~\ref{consecutiveroots})
are precisely the same (\ref{NumberRoots}) as for $Q(k,\alpha)$ 
(sec.~\ref{consecutiveroots}). (There strangely appears to be a sole exception to this rule for $\alpha=3$, where there are five such roots for $Q(k,3)$ and six such for $P(k,3)$, with $P(-3,3)$ anomalously equalling 0.) However, in the $P(k,\alpha)$ situation, the component polynomials are of degree $4 \alpha -2$, while in the $Q(k,\alpha)$ setting the corresponding polynomials are of considerably smaller degree  $\alpha +\left\lfloor \frac{\alpha -1}{2}\right\rfloor -1$, so we are faced with a greater number of coefficients to determine. 

Here, is the equation we have solved to determine--based on \cite[App. A]{LatestCollaboration}---formulas for 
$P(k,\alpha)$ for $\alpha=1,\ldots,47$. The c's are (nonnegative integer) coefficients we fitted to exact values obtained using the Legendre-polynomial density-approximation routine of 
Provost \cite{Provost}. (The first 15,761 of the moments (\ref{MomentFormula}) 
were employed.)
\begin{equation}
P(k,\alpha)= 1 - 
\end{equation}
\begin{displaymath}
 \resizebox{1.10\hsize}{!}{%
        $\frac{2^{8 \alpha +2 k+1} k^{-\left\lfloor \frac{\alpha +1}{3}\right\rfloor -3} \Gamma
   \left(k+3 \alpha +\frac{3}{2}\right) \Gamma (2 k+5 \alpha +2) \Gamma \left(k+3 \alpha
   +\left\lfloor \frac{\alpha +1}{3}\right\rfloor +1\right) \left(k^{\left\lfloor
   \frac{\alpha +1}{3}\right\rfloor +3} \left(\sum _{i=1}^{3 \alpha -\left\lfloor
   \frac{\alpha +1}{3}\right\rfloor +\left\lfloor \frac{\alpha +1}{2}\right\rfloor -3}
   c_{i+1} k^{i-1}\right)+\left(c_1+k\right) k^{3 \alpha +\left\lfloor \frac{\alpha
   +1}{2}\right\rfloor }\right)}{\sqrt{\pi } \Gamma (2 \alpha ) \Gamma (3 k+10 \alpha +2)
   \Gamma \left(k+2 \alpha +\left\lfloor \frac{\alpha +1}{2}\right\rfloor +1\right)}$%
	}.
\end{displaymath}
\subsection{The ratios $\frac{P(k+1,\alpha)-P(k,\alpha)}{Q(k+1,\alpha)-Q(k,\alpha)}$}
In  App.~\ref{AppTermByTermDifferencesFull}, we show  a number of formulas we have 
generated for the differences between the formulas for $P(k,\alpha)$ for successive
values of $k$, in relation to the earlier $Q(k,\alpha)$-based formulas shown in 
App.~\ref{TermByTermDifferences}. (A stark contrast occurs, with the formulas $k=-2,\dots,-10$ initially yielding [``biproper''] rational functions--with equal-degree numerators and [zero constant term] denominators [the degrees satisfying a certain difference equation]--and, then, difference equations for $k>-2$.)
So, it appears that the quest for a general $P(k,\alpha)$ formula could be successfully addressed by employing the same framework as in the $Q(k,\alpha)$ case, by modifying 
the $H(\alpha,j)$ function to incorporate the new terms shown in 
App.~\ref{TermByTermDifferences} and their extensions to $k$, in general. We see an evident relation between the coefficients of the $y[1+\alpha]$ terms in the difference equations in  App.~\ref{AppTermByTermDifferencesFull} and the six hypergeometric upper parameters described in sec.~\ref{UpperParameters} in the pattern of two 6's and four 5's.
Also, the coefficients of the $y[\alpha]$ terms appear related to the six hypergeometric lower parameters described in sec.~\ref{LowerParameters}. 

Further, in App.~\ref{AppTermByTermDifferencesFull2} we show the ratios as functions of 
$k$, rather than of $\alpha$. 
\subsubsection{Solution of difference equation for $\frac{P(1,\alpha)-P(0,\alpha)}{Q(1,\alpha)-Q(0,\alpha)}$} \label{PQratios}
We have been successfully able to solve the second difference equation recorded (in two forms) in App.~\ref{AppTermByTermDifferencesFull}. 
The initial solution consisted of a large
(multi-page) output with numerous hypergeometric functions (again with argument $\frac{27}{64}$). (In App.~\ref{CarlLove}, we show the Maple counterpart, provided by Carl Love 
(http://math.stackexchange.com/questions/1903720/what-solution-does-maple-give-to-this-difference-equation), of our Mathematica solution. There is an implicit [unperformed] summation in it.) The solution naturally broke into the sum of two parts. For the first part--using high-precision numerics, rationalizations and the FindSequenceFunction command--we were able to obtain the (hypergeometric-free) formula
\begin{equation} \label{firstpart}
\frac{5\ 3^{-3 \alpha -1} 8^{2 \alpha +1} (5 \alpha +3)
   \left(\frac{7}{10}\right)_{\alpha } \left(\frac{9}{10}\right)_{\alpha } (1)_{\alpha }
   \left(\frac{11}{10}\right)_{\alpha } \left(\frac{13}{10}\right)_{\alpha }
   \left(\frac{3}{2}\right)_{\alpha }}{(20 \alpha +11) \left(\frac{2}{5}\right)_{\alpha
   } \left(\frac{3}{5}\right)_{\alpha } \left(\frac{4}{5}\right)_{\alpha }
   \left(\frac{5}{6}\right)_{\alpha } \left(\frac{7}{6}\right)_{\alpha }
   \left(\frac{6}{5}\right)_{\alpha }}.
\end{equation}
Remarkably, when this term was multiplied by the function (which comprises the denominator of the ratio), examples of which are shown in App.~\ref{TermByTermDifferences}, and 
formulated in (\ref{Q(k+1,a)Formula2}),
\begin{equation}
Q(1,\alpha)-Q(0,\alpha) =\frac{\pi  2^{-4 \alpha } 3^{3 \alpha +1} 5^{-5 \alpha -3} (20 \alpha +11) \Gamma
   \left(\alpha +\frac{5}{6}\right) \Gamma \left(\alpha +\frac{7}{6}\right) \Gamma (5
   \alpha +2)}{\Gamma (\alpha ) \Gamma \left(\alpha +\frac{7}{10}\right) \Gamma
   \left(\alpha +\frac{9}{10}\right) \Gamma \left(\alpha +\frac{11}{10}\right) \Gamma
   \left(\alpha +\frac{13}{10}\right) \Gamma (2 \alpha +3)},
\end{equation}
the product simplified to the form $\frac{4 \alpha  (5 \alpha +3)}{9 (\alpha +1)}$. So, we  can consider this term to be the first of two parts of a formula for $P(1,\alpha)-P(0,\alpha)$. Now, in quest of the remaining term, when we formed a new difference equation for just the second part,  we obtained a new solution, again naturally breaking into the sum of two parts. Now, the first part--previously given by (\ref{firstpart})--was zero, and the new second part was given by precisely the same difference equation as originally, but for the single change of the initial value (at $\alpha=1$) from $y[1]=\frac{158}{31} = \frac{474}{93}$ to $y[1]=-\frac{4102}{93}$.
\subsection{$X$-states counterpart}
In App.~\ref{AppendixXstates} we show the analogue of the 
$P(k,\alpha)$ formulas for the ``toy'' model of $X$-states \cite{LatestCollaboration2,Xstates2}. One feature to be immediately noted is that the 
arguments of the indicated hypergeometric functions are -1. Another is that for half-integer  $\alpha$'s, $P(k,\alpha)$ yields rational values, while $P_{X-states}(k,\alpha)$
yields value of the form 1 minus rational numbers divided by $\pi^2$.
\subsection{Use of consecutive negative roots}
We have noted that both $Q(k,\alpha)$ and $P(k,\alpha)$ have roots at consecutive negative values of $k$ (sec.~\ref{consecutiveroots}). If we examine the (limiting) values of 
$P(k,\alpha)$ for $k$ immediately (one) below the end of the consecutive series, we find that they satisfy the relation
\begin{equation} \label{StartingPoint}
P(-\frac{1}{4} (-1)^{\alpha } \left((-1)^{\alpha } (10 \alpha +1)-1\right),\alpha)=\frac{(3 \alpha  (5 \alpha +2)-1) \sin \left(\frac{\pi  \alpha }{2}\right)}{4 (\alpha
   +1)}+\cos \left(\frac{\pi  \alpha }{2}\right).
\end{equation}
(This might serve as a "starting point" analogous to the use ((\ref{CharlesInteger}),  (\ref{Qaa})) of $Q(-\alpha,\alpha)$). For the analogous set of $Q(-\frac{1}{4} (-1)^{\alpha } \left((-1)^{\alpha } (10 \alpha +1)-1\right),\alpha)$'s, the real parts appear to be $\frac{1}{2}$ for even $\alpha$ and $-\frac{1}{4}$ for odd $\alpha$, with the imaginary parts given by 
\begin{equation}
\Im{Q(-\frac{1}{4} (-1)^{\alpha } \left((-1)^{\alpha } (10 \alpha +1)-1\right),\alpha)}=-\frac{3 (-1)^{\alpha } \left(20 \left((-1)^{\alpha }+3\right) \alpha +5 (-1)^{\alpha
   }+7\right)}{4 \pi  \left(400 \alpha ^2+80 \alpha +3\right)}.
\end{equation}

Dunkl has observed that the sequence generated by 
(\ref{StartingPoint})  is really two interspersed sequences, one for odd and one for even values of $\alpha$.
They can be represented as $f(2 \alpha)=(-1)^{\alpha}$ and $f(2 \alpha+1) =\frac{(-1)^{\alpha } \left(15 \alpha ^2+18 \alpha +5\right)}{2 \alpha +2} = (-1)^{\alpha } \left(\frac{15 \alpha }{2}+\frac{1}{\alpha +1}+\frac{3}{2}\right).$
\subsection{Setting $k$ so that the $_6F_5$ parameters in the $Q(k,\alpha)$ formula are zero}
It appeared to be an exercise of interest to set $k$ in $P(k,\alpha)$ so that, in turn, one of the five variable upper and lower parameters in the $_6F_5$ function in the formula   for $Q(k,\alpha)$ would equal zero. We now enumerate those such scenarios, for which we were able to construct formulas.

For $k \rightarrow -2 - 4 \alpha$, we found that 
\begin{equation}
P(k,\alpha)= \frac{3\ 2^{4 \alpha -1} (5 \alpha +2) \Gamma \left(2 \alpha
   +\frac{3}{2}\right)}{\sqrt{\pi } (3 \alpha +2) \Gamma (2 \alpha +2)}+\frac{1}{4}.
\end{equation} 

As already observed, since we have consecutive roots descending downward from $-1-\alpha$, for $k=-2 -\alpha$, we have $P(k,\alpha)=0$.

Further, we found that, in the limit $k \rightarrow -1-\frac{5 \alpha}{2}$,
\begin{align*}
P(k,\alpha) = -\infty, \hspace{.1cm} \alpha &\equiv 1 \mod 4 \\
P(k,\alpha) = -1, \hspace{.1cm} \alpha &\equiv 2 \mod 4 \\
P(k,\alpha) = \infty, \hspace{.1cm} \alpha &\equiv 3 \mod 4 \\
P(k,\alpha) = 1, \hspace{.1cm} \alpha &\equiv 0 \mod 4 \\.
\end{align*}
Also, for $k \rightarrow -\frac{3}{2}-\frac{5 \alpha}{2}$, for  even $\alpha$
\begin{align*}
P(k,\alpha) = -\infty, \hspace{.1cm} \alpha &\equiv 2 \mod 4 \\
P(k,\alpha) = \infty, \hspace{.1cm} \alpha &\equiv 0 \mod 4 \\,
\end{align*}
and for odd $\alpha$
\begin{align*}
P(k,\alpha) =-\frac{i i^{\alpha } (3 \alpha  (5 \alpha +2)-1)}{4 (\alpha +1)}.
\end{align*} 

Additionally, along similar investigative lines, we have the  $P(k,\alpha)$ formulas in  App.~\ref{P(f(a),a)}--particularly elegantly 
($k \rightarrow -1 - 4 \alpha$),
\begin{equation}
P(k,\alpha) = \frac{1}{4} \left(\frac{3\ 16^{\alpha } \Gamma \left(2 \alpha
   +\frac{1}{2}\right)}{\sqrt{\pi } \Gamma (2 \alpha +1)}+1\right).
\end{equation} 

\subsection{Formula for $P(-\alpha,\alpha)$}
Also, eventually (after having computed $P(k,\alpha)$ for $\alpha=1,\ldots,49$), we were able to obtain the formula (not as explicit as that for $Q(-\alpha,\alpha)$) shown in App.~\ref{fig:P(-a,a)} for $P(-\alpha,\alpha)$.

\subsection{Two $P(k,\alpha)$ formulas involving the Lerch transcendent}
In the limit $k-\rightarrow -\frac{1}{2} -\frac{5 \alpha}{2}$, we found for {\it even} $\alpha$,
\begin{equation}
P(k,\alpha)=
\end{equation}
\begin{displaymath}
\resizebox{1.10\hsize}{!}{%
        $\frac{i^{\alpha } \left(3 \Phi \left(-1,1,\frac{\alpha }{2}+\frac{1}{10}\right)+5 \Phi
   \left(-1,1,\frac{\alpha }{2}+\frac{1}{6}\right)-3 \Phi \left(-1,1,\frac{\alpha
   }{2}+\frac{3}{10}\right)-3 \Phi \left(-1,1,\frac{\alpha }{2}+\frac{7}{10}\right)+5
   \Phi \left(-1,1,\frac{\alpha }{2}+\frac{5}{6}\right)+3 \Phi \left(-1,1,\frac{\alpha
   }{2}+\frac{9}{10}\right)-2 \Phi \left(-1,1,\frac{\alpha +1}{2}\right)\right)}{15 \pi }$%
	}.
\end{displaymath}
Here, the Lerch transcendant $\Phi (z,s,b)=\Sigma_{i=0}^{\infty} z^i/(i+b)^s$.
In the same limit, we have for {\it odd} $\alpha$,
\begin{equation}
Q(k,\alpha)=\frac{3 \sqrt{\frac{\pi }{5}} \Gamma \left(\frac{\alpha }{2}+\frac{7}{10}\right) \Gamma
   \left(\frac{\alpha }{2}+\frac{9}{10}\right) \Gamma \left(\frac{\alpha
   }{2}+\frac{11}{10}\right) \Gamma \left(\frac{\alpha }{2}+\frac{13}{10}\right) \Gamma
   \left(\frac{\alpha +1}{2}\right)}{8 \Gamma \left(\frac{\alpha }{2}+\frac{3}{5}\right)
   \Gamma \left(\frac{\alpha }{2}+\frac{4}{5}\right) \Gamma \left(\frac{\alpha
   }{2}+1\right) \Gamma \left(\frac{\alpha }{2}+\frac{6}{5}\right) \Gamma
   \left(\frac{\alpha }{2}+\frac{7}{5}\right)}.
\end{equation}

Next, in the limit $k-\rightarrow  -\frac{5 \alpha}{2}$, we found for {\it odd} $\alpha$,
\begin{equation}
P(k,\alpha)=
\end{equation}
\begin{displaymath}
\resizebox{1.10\hsize}{!}{%
        $-\frac{i e^{\frac{i \pi  \alpha }{2}} \left(-40 \alpha +(5 \alpha +2) \alpha  \left(5
   \Phi \left(-1,1,\frac{\alpha }{2}+\frac{1}{3}\right)+3 \Phi \left(-1,1,\frac{\alpha
   }{2}+\frac{2}{5}\right)-3 \Phi \left(-1,1,\frac{\alpha }{2}+\frac{3}{5}\right)-5 \Phi
   \left(-1,1,\frac{\alpha }{2}+\frac{2}{3}\right)+3 \Phi \left(-1,1,\frac{\alpha
   }{2}+\frac{4}{5}\right)+2 \Phi \left(-1,1,\frac{\alpha }{2}+1\right)+3 \Phi
   \left(-1,1,\frac{\alpha }{2}+\frac{6}{5}\right)\right)+12\right)}{15 \pi  \alpha  (5
   \alpha +2)}$
	},
\end{displaymath}
while for {\it even} $\alpha$,
\begin{equation}
Q(k,\alpha)=\frac{3 \pi ^{5/2} 5^{-\frac{5 \alpha }{2}-2} (35 \alpha +22) \Gamma \left(\frac{5 \alpha
   }{2}+2\right)}{88 \Gamma \left(\frac{\alpha }{2}+\frac{7}{10}\right) \Gamma
   \left(\frac{\alpha }{2}+\frac{9}{10}\right) \Gamma \left(\frac{\alpha
   }{2}+\frac{11}{10}\right) \Gamma \left(\frac{\alpha }{2}+\frac{13}{10}\right) \Gamma
   \left(\frac{\alpha +3}{2}\right)}.
\end{equation}
We note the similarities in integer coefficients between the two Lerch-based $P(k,\alpha)$ formulas, and the by now familiar occurrences (secs.~\ref{First7F6},\ref{PQratios}, 
App.~\ref{fig:kminus1-kplus2}) of simple 
fractions with denominators that are multiples of five and six.

\subsection{Rules for leading coefficients of the polynomials $p_{\alpha}(k)$}
In App.~\ref{CoefficientRules} we show for $i=1,\ldots,10$, the first of the rules we have developed for the leading coefficients of the polynomials $p_{\alpha}(k)$ given in 
the formula above (\ref{HindawiFormula}) for $P(k,\alpha)$--having been normalized to monic form
(the original leading degree-($4 \alpha-2$) coefficient being $\frac{2^{8 \alpha+1}}{2 \alpha-1)!}$). (For convenience, we drop this $k^{4 \alpha -2}$ term, and are left with 
degree-($4 \alpha-3)$ polynomials.) We note that these resultant polynomials are of degree $2 i$.
Now, we can make the interesting observation (essentially putting the polynomial in Horner form) that their leading (highest power) coefficients are given (in descending order)
by the rules:
\begin{equation}
C_1=\frac{\left(\frac{17}{2}\right)^i}{\Gamma (i+1)},
\end{equation}
\begin{equation}
C_2=\frac{2^{-i-2} 17^{i-2} (1109-497 i)}{3 \Gamma (i)},
\end{equation}
\begin{equation}
C_3=\frac{2^{-i-5} 17^{i-4} (i (i (247009 i-1370262)+3942323)-11308734)}{9 \Gamma (i)}.
\end{equation}
Also, $C_4$ is the product of 
\begin{equation}
-\frac{2^{-i-7} 17^{i-6} (i-1) i}{405 \Gamma (i+1)}
\end{equation}
and
\begin{equation}
613817365 i^4-5492491130 i^3+30016283027 i^2-173872269670 i+542508998592.
\end{equation}Further, $C_5$ is the product of 
\begin{equation}
\frac{2^{-i-11} 17^{i-8} (i-1) i}{1215 \Gamma (i+1)}
\end{equation} 
and 
\begin{equation}
305067230405 i^6-4403156498055 i^5+38051293414691 i^4
\end{equation}
\begin{displaymath}
-325978342903557
   i^3+2137571940201488 i^2-8722204904328012 i+13657232612174832.
\end{displaymath}
Continuing, $C_6$ is the product of 
\begin{equation}
-\frac{2^{-i-13} 17^{i-10} (i-2) (i-1) i}{25515 \Gamma (i+1)}
\end{equation}
and
\begin{displaymath}
212265778915799 i^7-4033760477145378 i^6+46257531538470350 i^5-526319720165886192
   i^4
\end{displaymath}
\begin{displaymath}
+5002806671861237555 i^3-35895786322816308558 i^2+169446873953910154824
   i-385892347895176978944,
\end{displaymath}
while $C_7$ is the product of 
\begin{equation}
\frac{2^{-i-16} 17^{i-12} (i-2) (i-1) i}{1148175 \Gamma (i+1)}
\end{equation}
and
\begin{displaymath}
527480460605760515 i^9-14061542253335879085 i^8+216128338841103270330
   i^7-3070915881213672409050 i^6
\end{displaymath}
\begin{displaymath}
+39074939804872696010811 i^5-414647891239558549971645
   i^4+3466800379462987766973880 i^3
\end{displaymath}
\begin{displaymath}
-20874814527662001270399420
   i^2+78054176824402526959936464 i-118165465673929410155118720.
\end{displaymath}
So, an obvious important challenge would be to find the common formula
generating these results. (The pattern of [negative] integer exponents of 2--that is, 
0,2,5,7,11,13,16--is yielded by sequence A004134 "Denominators in expansion of $(1-x)^{-1/4}$ are $2^a(n)$" of the The On-Line Encyclopedia of Integer Sequences.)

Let us make the observation that the constant (lowest-order) coefficient in the polynomial $p_k(\alpha)$ in the formula for $P(k,\alpha)$ in (\ref{HindawiFormula}) is equal to
$\frac{1-2 Q(0,\alpha)}{G(0,\alpha)}$.
\section{Concluding Remarks}
The asymptotic analyses reported here and those in studies of Szarek, Aubrun and Ye \cite{sz1,aubrun2,sz2}  
both employ
Hilbert-Schmidt and (more generally) random induced measures (cf. \cite{HitandRun}). However, contrastingly, we chiefly consider asymptotics as the
Dyson-index-like parameter $\alpha \rightarrow \infty$ (cf. \cite{MatrixModels,DUMITRIU20051083}), while they implicitly are concerned with the standard case of $\alpha=1$, and large numbers of qubits. Perhaps some relation exists, however, between
their high-dimensional findings and the quite limited set of asymptotics we have presented above (secs.~\ref{kAsymptotics1}, \ref{kAsymptotics2}, \ref{kAsymptotics3}), pertaining to the dimensional index $k \rightarrow \infty$.

A strong, intriguing theme in the analyses presented above has been the repeated occurrence of the interesting
constant  $z=\frac{27}{64} =(\frac{3}{4})^3$.  Let us note that J. Guillera in his article ``A new Ramanujan-like series for $\frac{1}{\pi^2}$'', applying methods related
to Zeilberger's algorithm \cite{doron}, obtained a hypergeometric identity involving a sum over $n$ from 0 to $\infty$ of terms involving factors of the form $(\frac{27}{64})^n$ 
\cite[sec. 3]{guillera} (cf. \cite[sec. 8]{chu2014accelerating}). 

Further, in a study of products of Ginibre
matrices of Penson and {\.Z}yczkowski, the Fuss-Catalan distribution $P_s(x)$ is represented as a sum of $s$ generalized hypergeometric functions $_{s}F_{s-1}$, somewhat analogous to those given above in Figs. 3-6 (and, in particular, Fig. 3 in \cite{slaterJModPhys}, since only $_{7}F_{6}$ functions are employed). These functions $P_s(x)$ have hypergeometric arguments 
$\frac{s^s}{(s+1)^{(s+1)}} x$, where $s$ is a nonnegative integer, and have support $x \in [0,\frac{s^s}{(s+1)^{(s+1)}}]$ \cite[eq. (11)]{karolkarol}. So, for $s=3$, $\frac{s^s}{(s+1)^{(s+1)}} =\frac{27}{256}$. (We had inquired of Hou whether the telescoping procedure might be profitably applied in such a context. He replied ``the method I used only works for $_{s}F_{s-1}$ with a concrete integer $s$'' [cf. \cite[eqs. (13)-(16)]{karolkarol}].) 
As an item of further curiosity, we note that in the MathWorld entry on hypergeometric functions, the identity $\, _2F_1\left(\frac{1}{3},\frac{2}{3};\frac{5}{6};\frac{27}{32}\right) =\frac{8}{5}$, the argument being $\frac{27}{32}$, is noted. (Also, cf. 
(\ref{Qaa}) above.)

\appendix
\section{Hypergeometric forms of the factors $G_2^k(\alpha)$} \label{fig:kminus1-kplus2}
\includegraphics[page=1,scale=0.9]{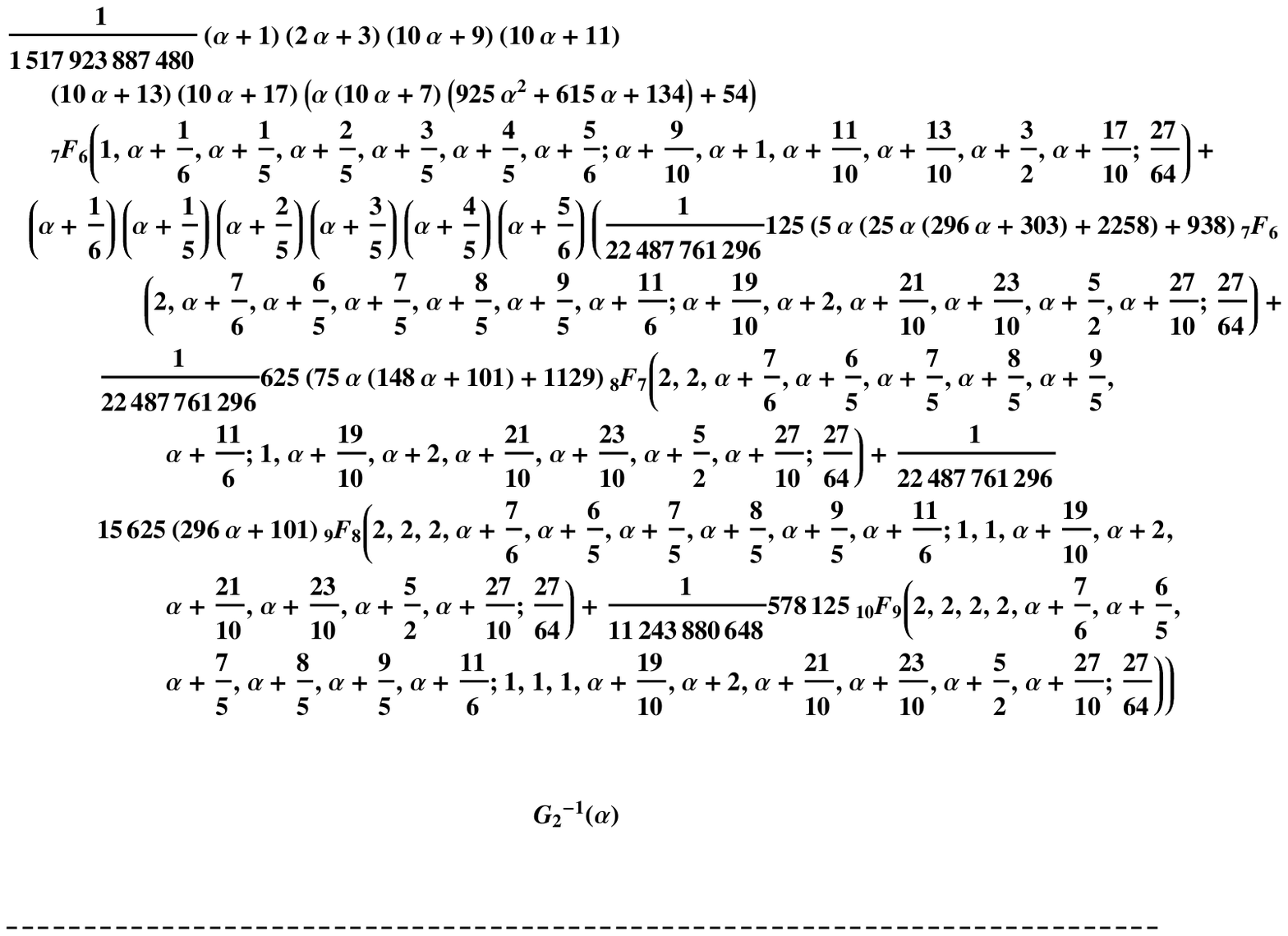}
\includegraphics[page=2,scale=0.9]{G2Formulas.pdf}
\includegraphics[page=3,scale=0.9]{G2Formulas.pdf}
\includegraphics[page=4,scale=0.9]{G2Formulas.pdf}
\section{Difference equation forms of the factors $G_2^k(\alpha)$} \label{DiffEqs}
\includegraphics[page=1,scale=0.9]{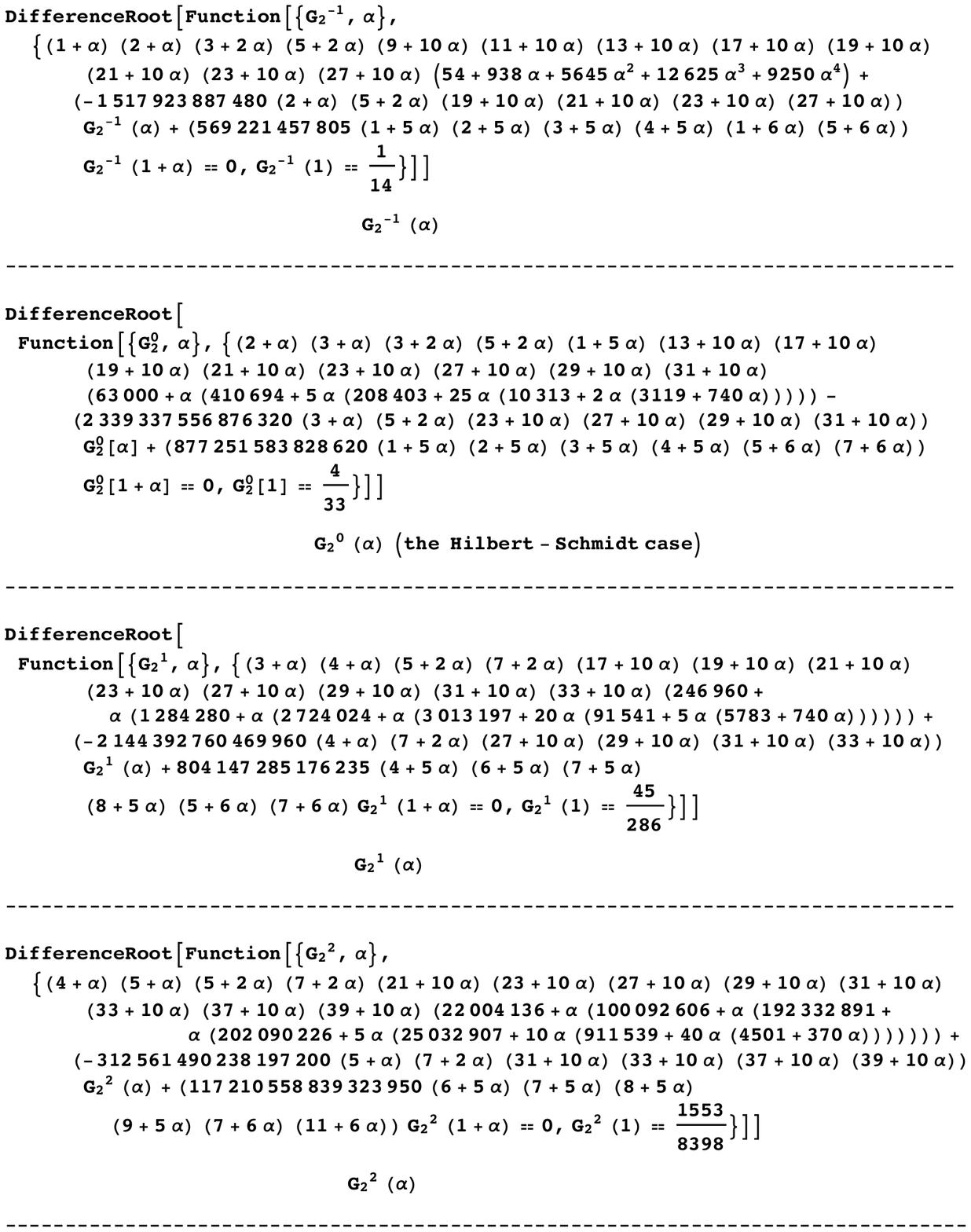}
\includegraphics[page=1,scale=0.9]{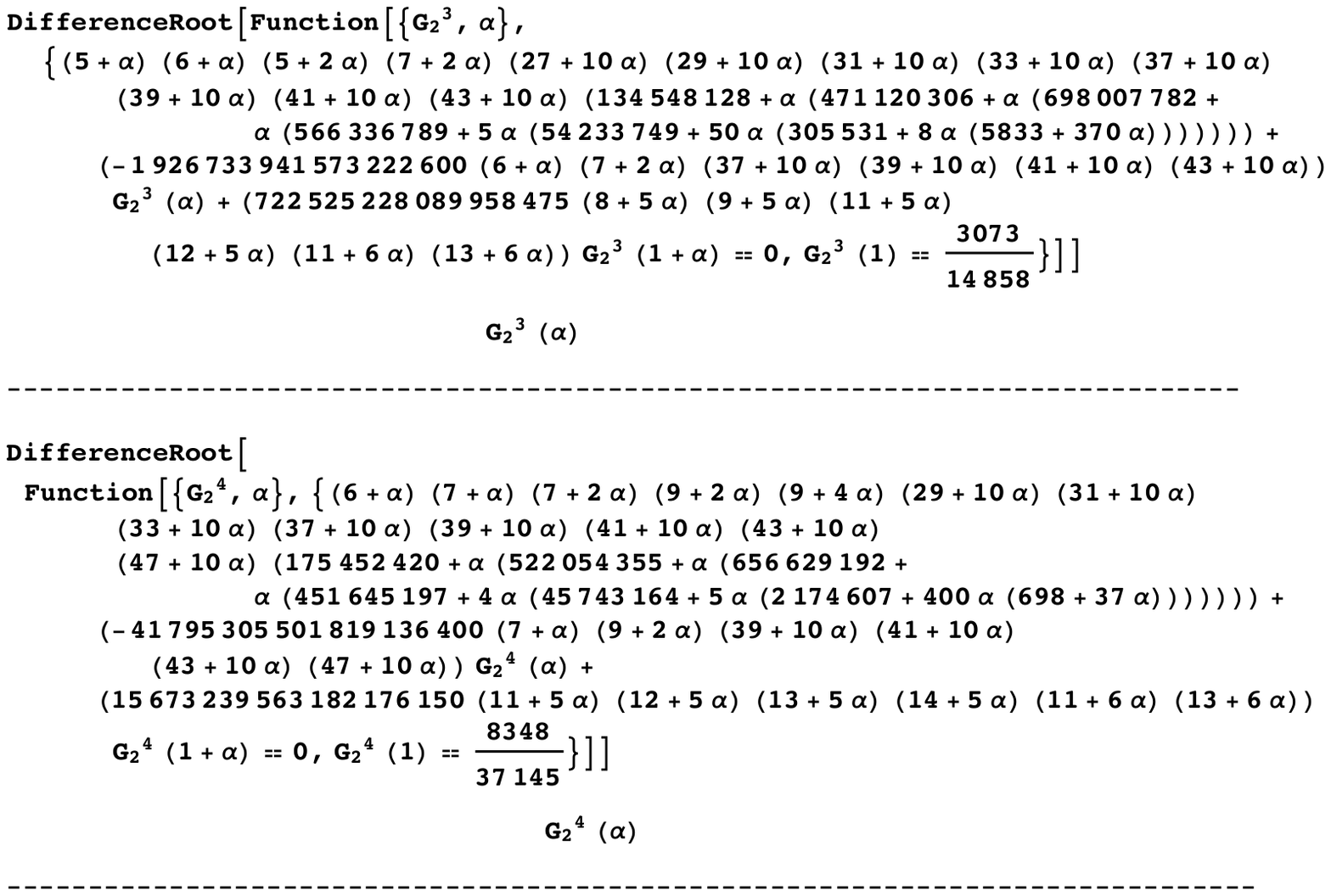}
\section{Hypergeometric-Free Formulas for 
$Q(k+1,\alpha)-Q(k,\alpha)$} \label{TermByTermDifferences}
\includegraphics[page=1,scale=0.9]{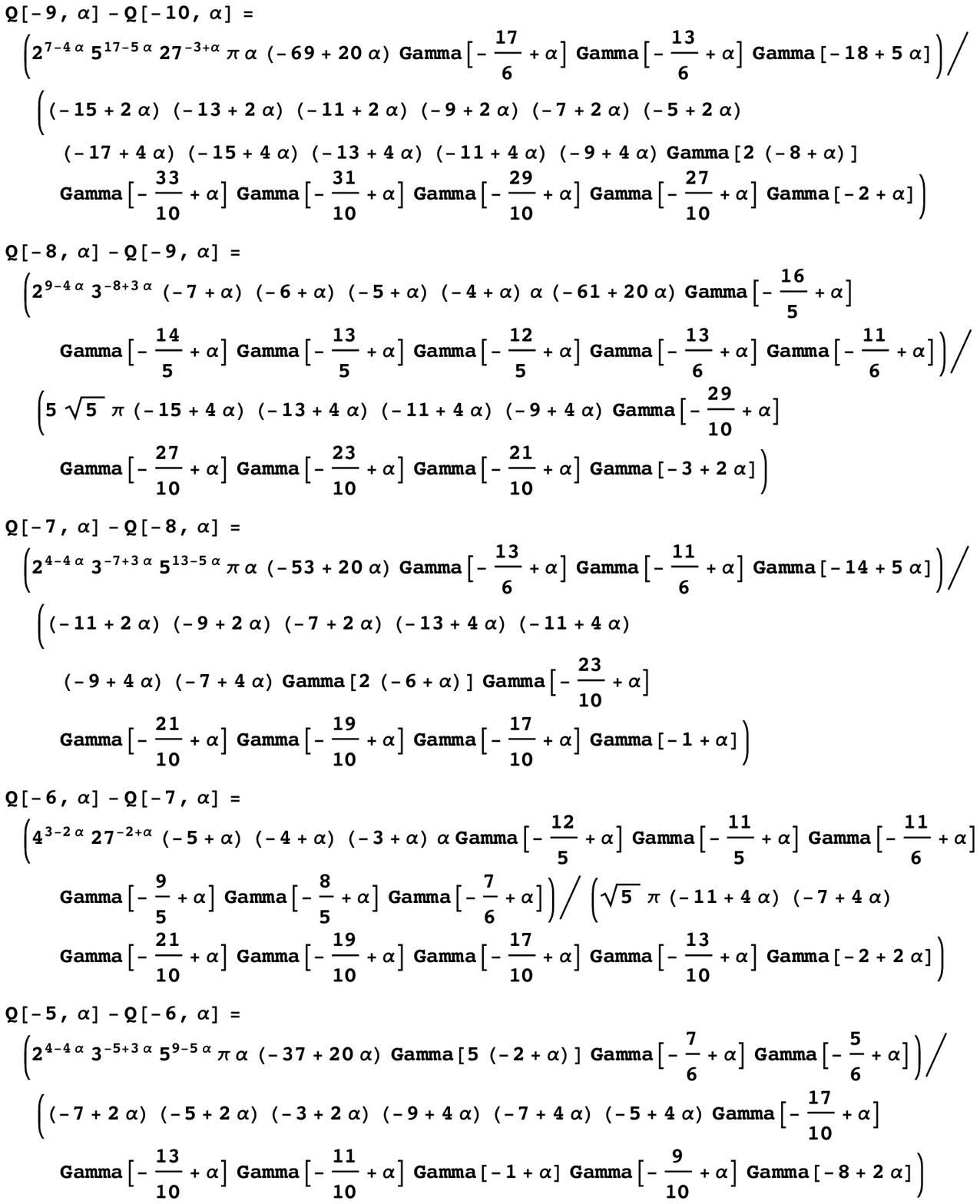}
\includegraphics[page=2,scale=0.9]{TermByTermDifferences.pdf}
\includegraphics[page=3,scale=0.9]{TermByTermDifferences.pdf}
\section{Maple worksheet of Qing-Hu Hou for $Q(1,\alpha)$ ``concise'' formula (\ref{k1Hou1})} \label{AppendixHou}
\includegraphics[page=1,scale=.85]{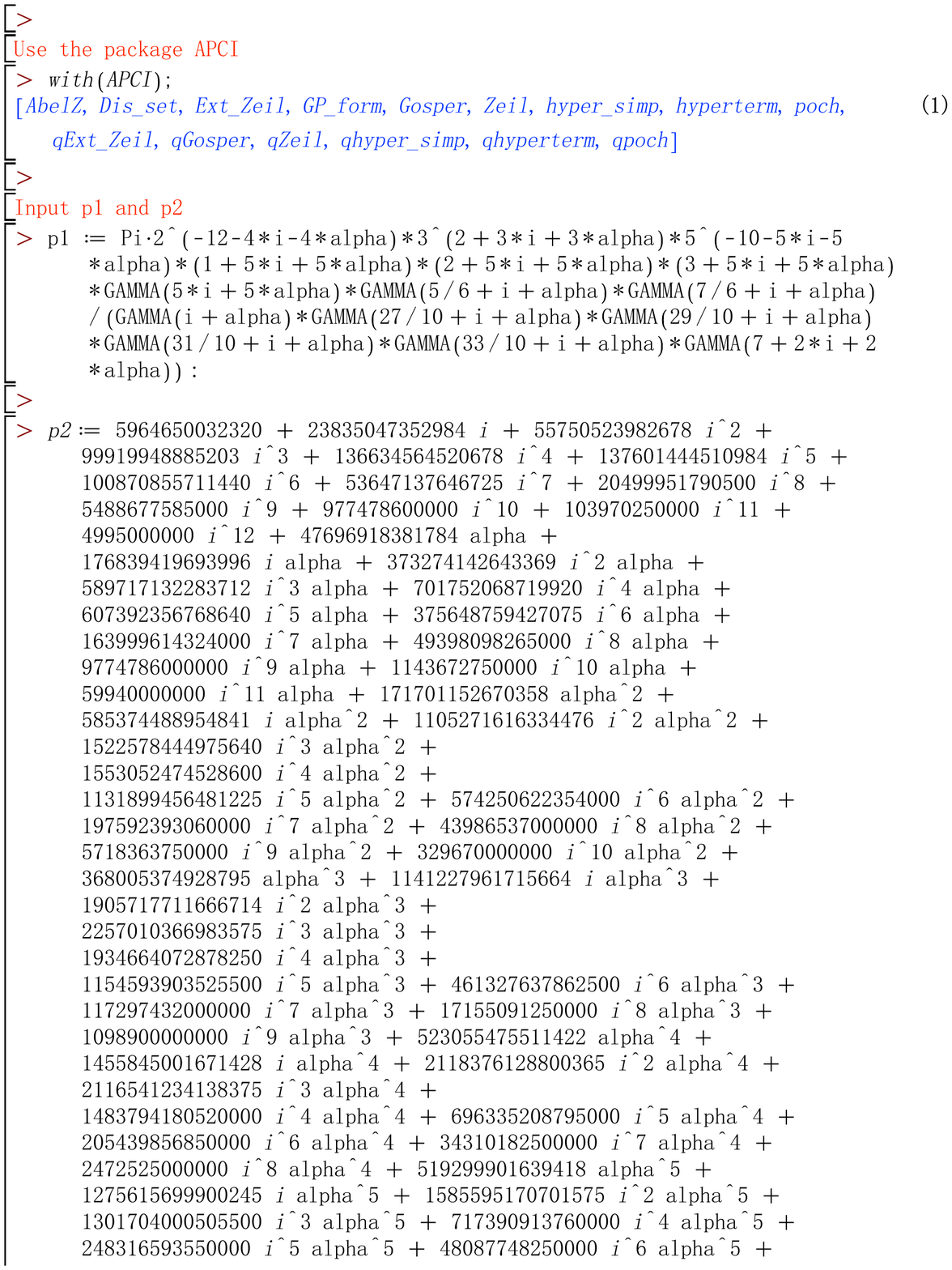}
\includegraphics[page=2,scale=.85]{k1Maple0.pdf}
\includegraphics[page=3,scale=.85]{k1Maple0.pdf}
\includegraphics[page=4,scale=.85]{k1Maple0.pdf} 
\section{Collected $Q(k,\alpha)$ formulas} \label{AppendixFormulas}
\includegraphics[page=1,scale=.95]{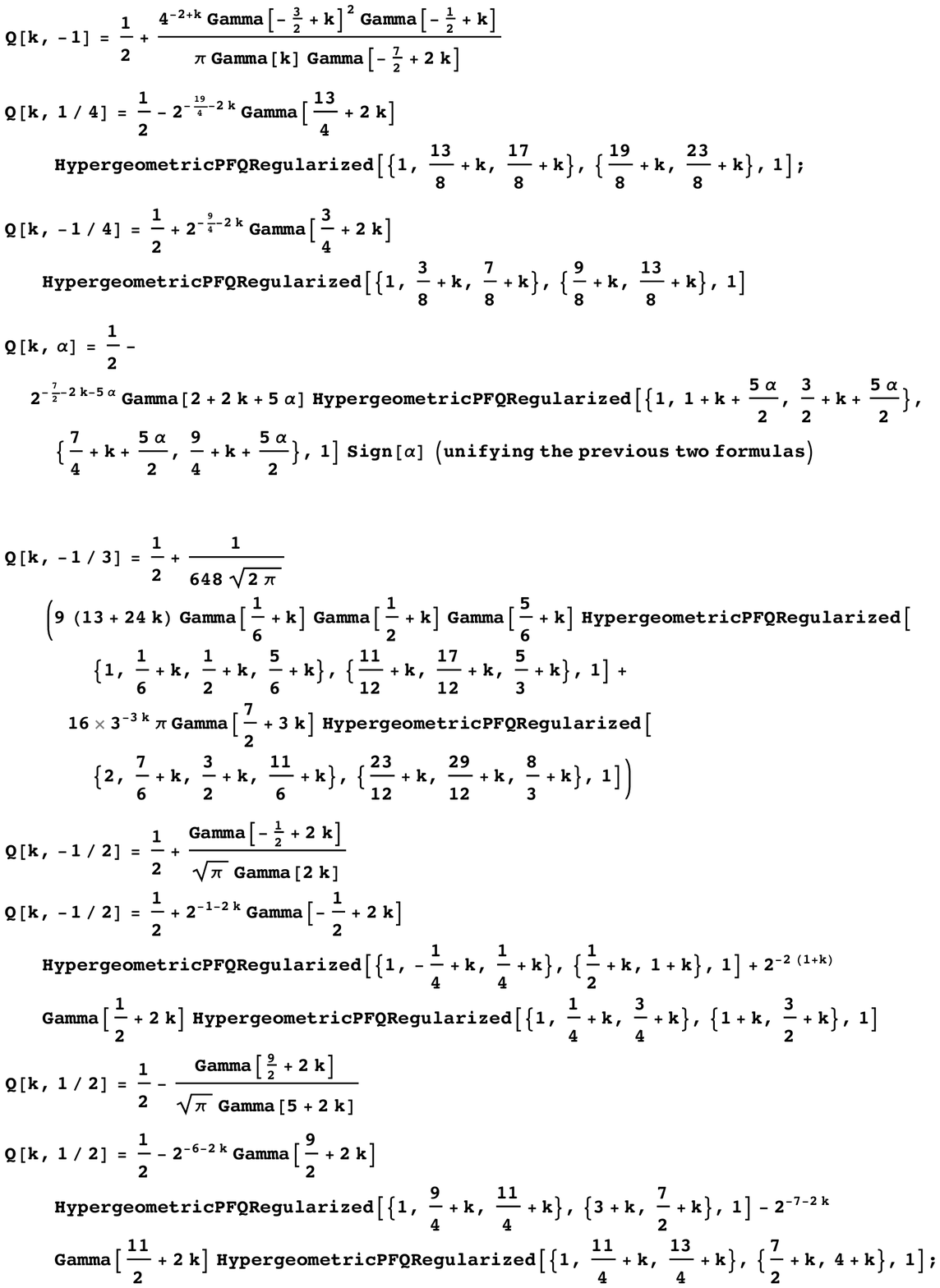}
\includegraphics[page=2,scale=.95]{FormulaSummary.pdf}
\includegraphics[page=3,scale=.95]{FormulaSummary.pdf}
\includegraphics[page=4,scale=.95]{FormulaSummary.pdf} 
\section{Collected $P(k,\alpha)$ formulas} \label{AppendixFormulasTotalProbs}
\includegraphics[page=1,scale=.9]{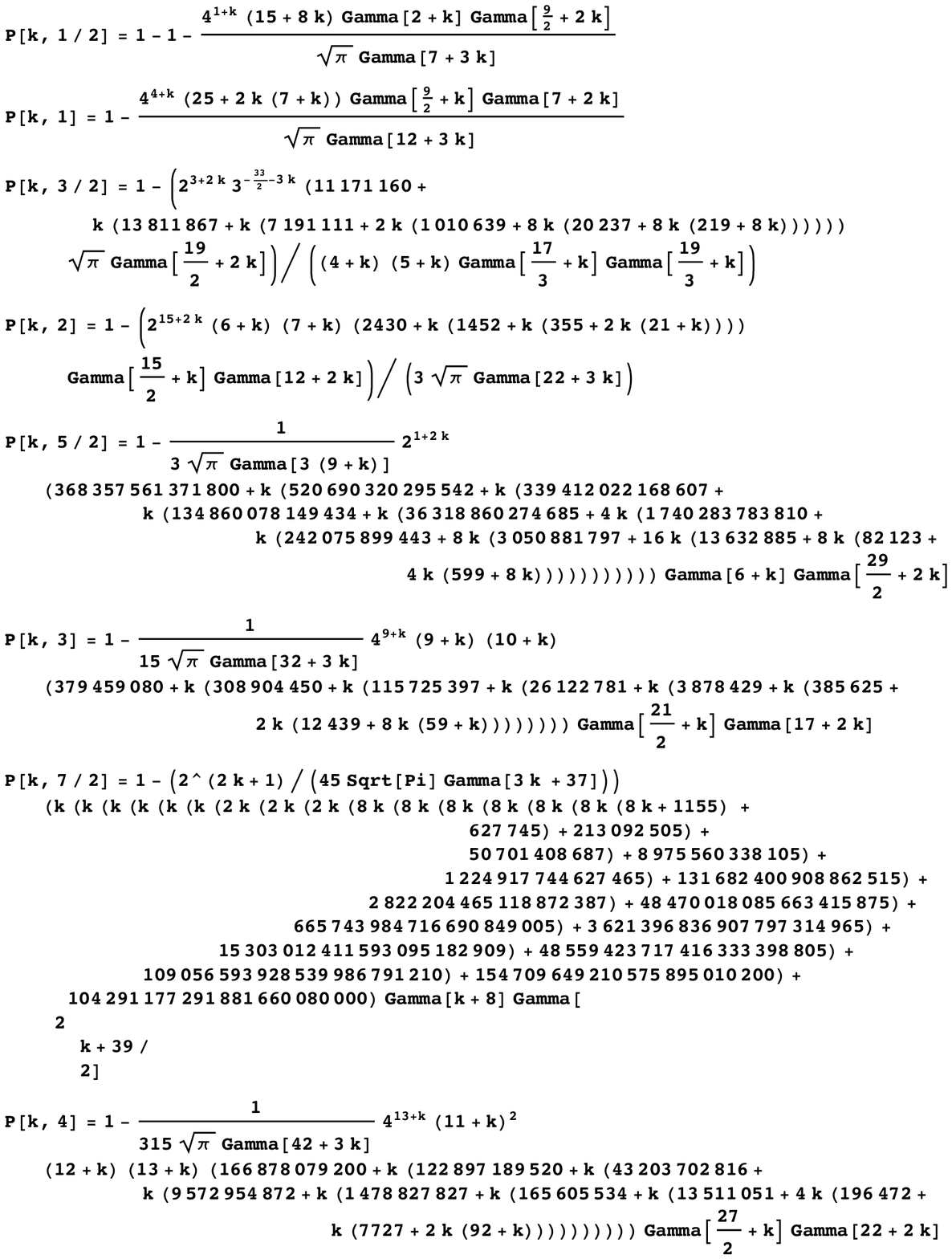}
\includegraphics[page=2,scale=.9]{FormulaSummaryTotalProbs.pdf}
\includegraphics[page=3,scale=.95]{FormulaSummaryTotalProbs.pdf}
\includegraphics[page=4,scale=.95]{FormulaSummaryTotalProbs.pdf}
\includegraphics[page=5,scale=.95]{FormulaSummaryTotalProbs.pdf}
\section{Formulas for the ratios $\frac{P(k+1,\alpha)-P(k,\alpha)}{Q(k+1,\alpha)-Q(k,\alpha)}$ as functions of $\alpha$} \label{AppTermByTermDifferencesFull}
\includegraphics[page=1,scale=0.9]{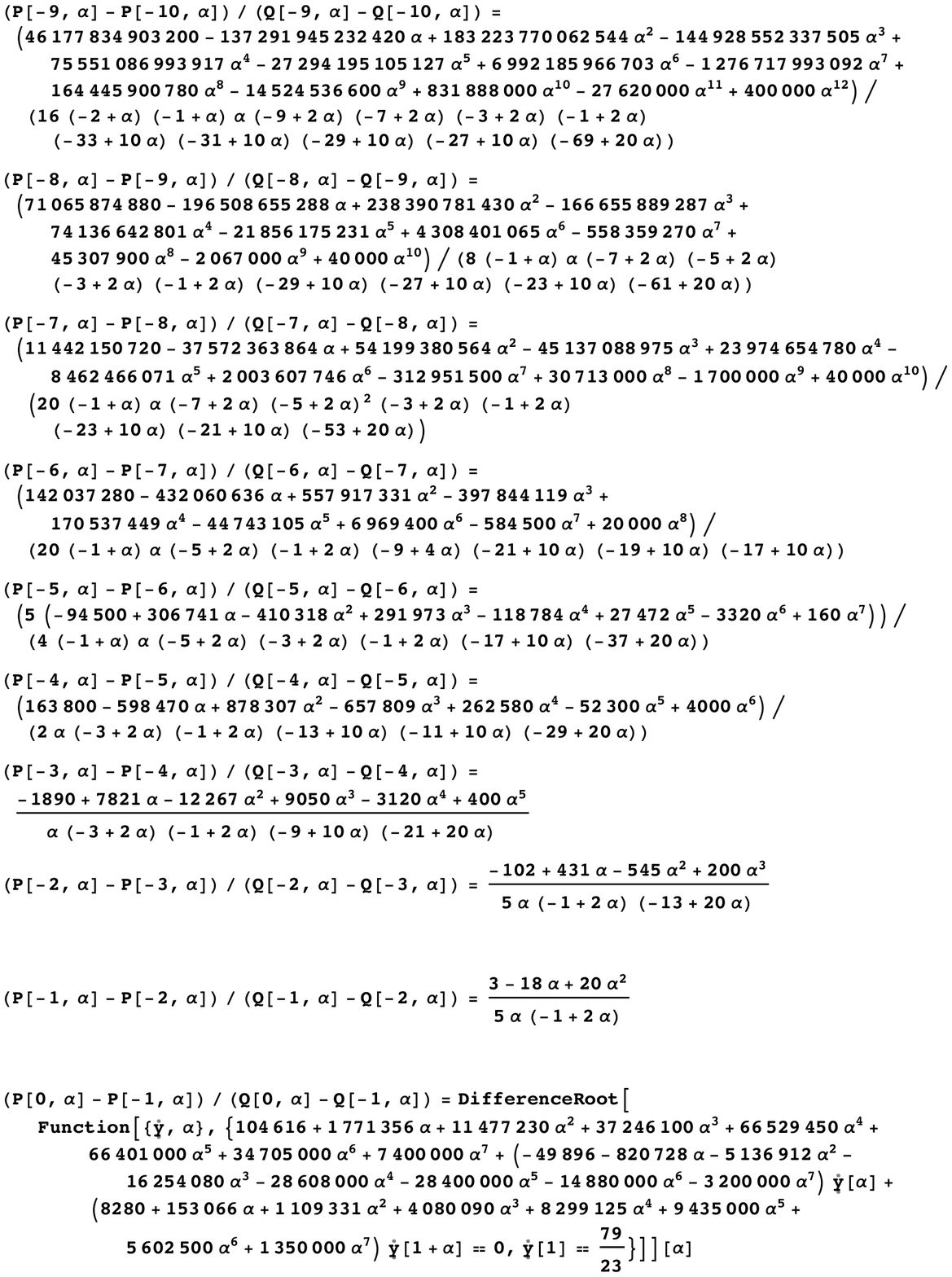}
\includegraphics[page=2,scale=0.9]{TermByTermDifferencesFull3.pdf}
\includegraphics[page=3,scale=0.9]{TermByTermDifferencesFull3.pdf}
\includegraphics[page=4,scale=0.9]{TermByTermDifferencesFull3.pdf}
\includegraphics[page=5,scale=0.9]{TermByTermDifferencesFull3.pdf}
\section{Formulas for the ratios $\frac{P(k+1,\alpha)-P(k,\alpha)}{Q(k+1,\alpha)-Q(k,\alpha)}$ as functions of $k$} \label{AppTermByTermDifferencesFull2}
\includegraphics[page=1,scale=0.9]{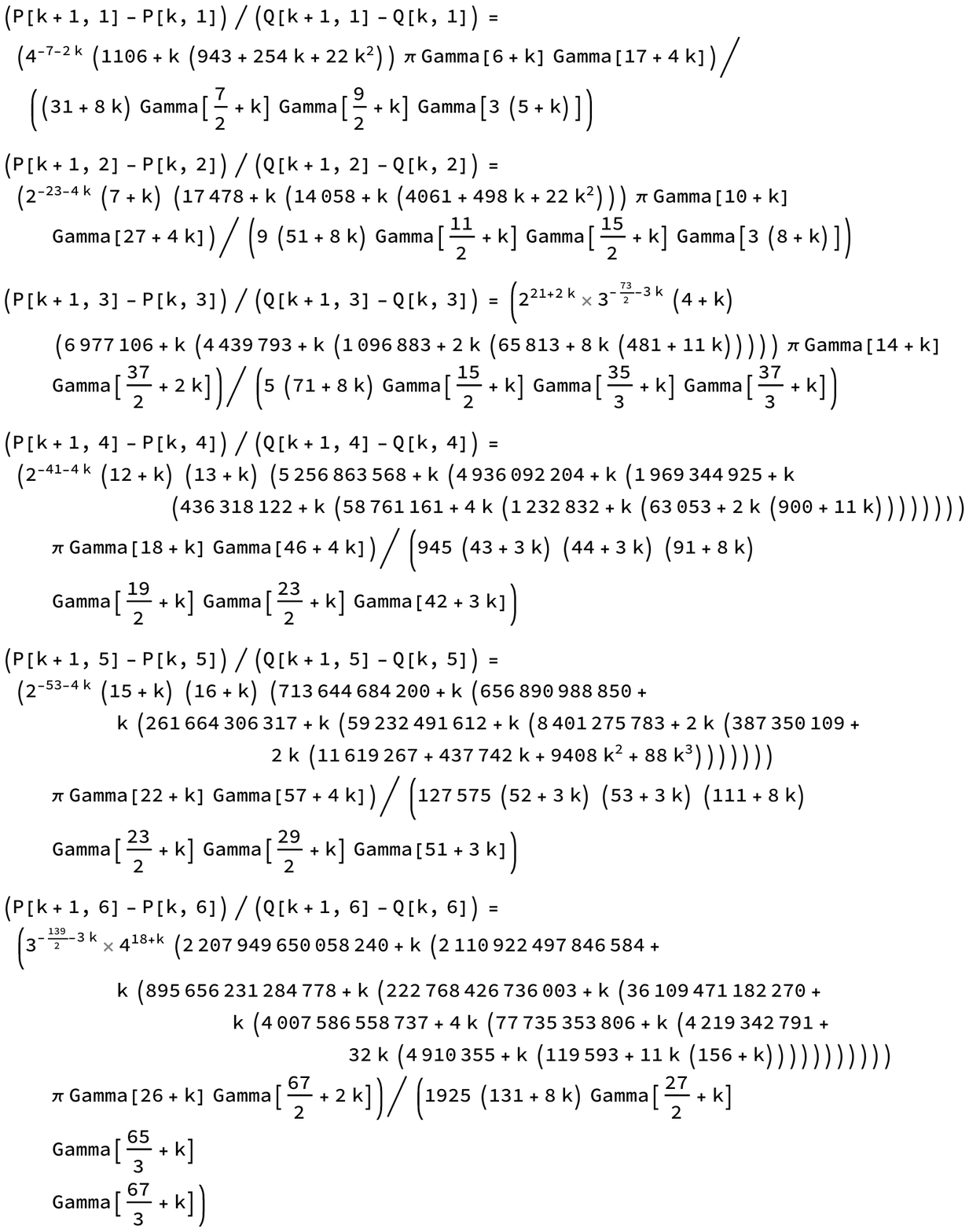}
\includegraphics[page=2,scale=0.9]{TermByTermDifferencesAlpha2.pdf}
\section{Maple solution, provided by Carl Love, of difference equation (App.~\ref{AppTermByTermDifferencesFull2}) for
$\frac{P(1,\alpha)-P(0,\alpha)}{Q(1,\alpha)-Q(0,\alpha)}$} \label{CarlLove}
\includegraphics[page=1,scale=0.85]{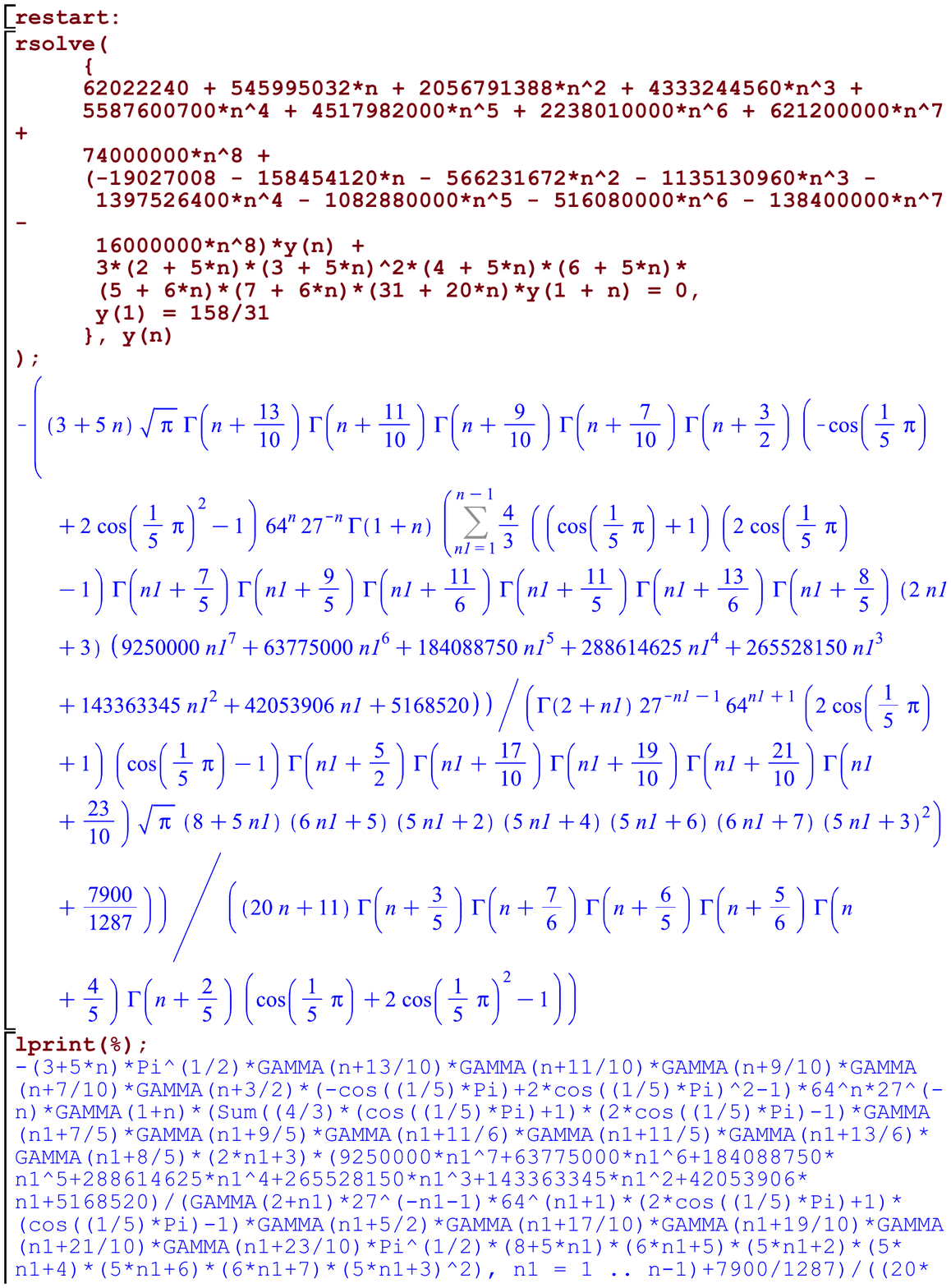}
\includegraphics[page=2,scale=0.85]{CarlLove.pdf}
\section{$P_{X-states}(k,\alpha)$ formulas} \label{AppendixXstates}
\includegraphics[page=1,scale=.9]{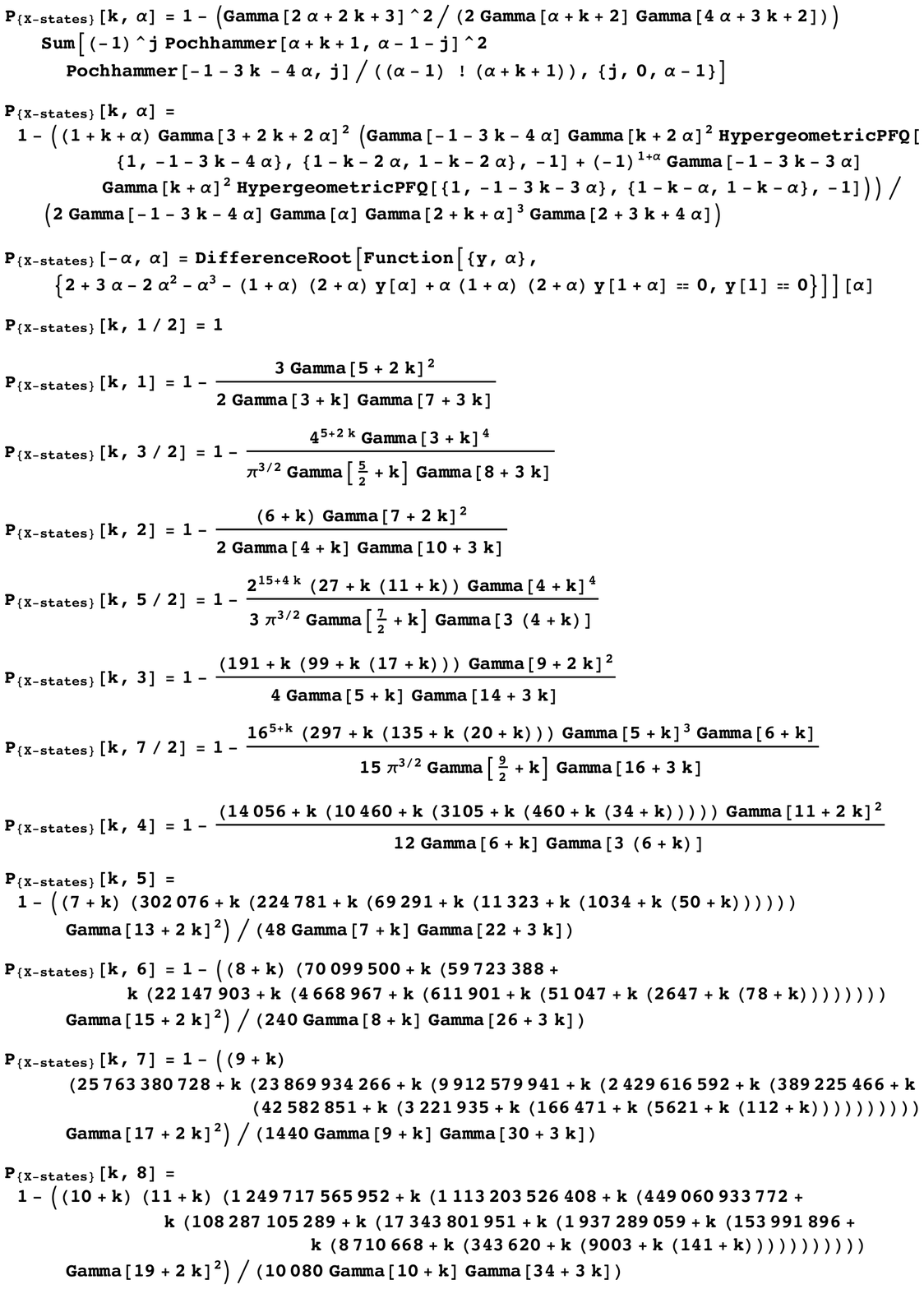}
\includegraphics[page=2,scale=.9]{FormulaSummaryXstates.pdf}
\section{Formula for $P(-1,\alpha)$}\label{fig:P(-1,a)}
\includegraphics[scale=0.95]{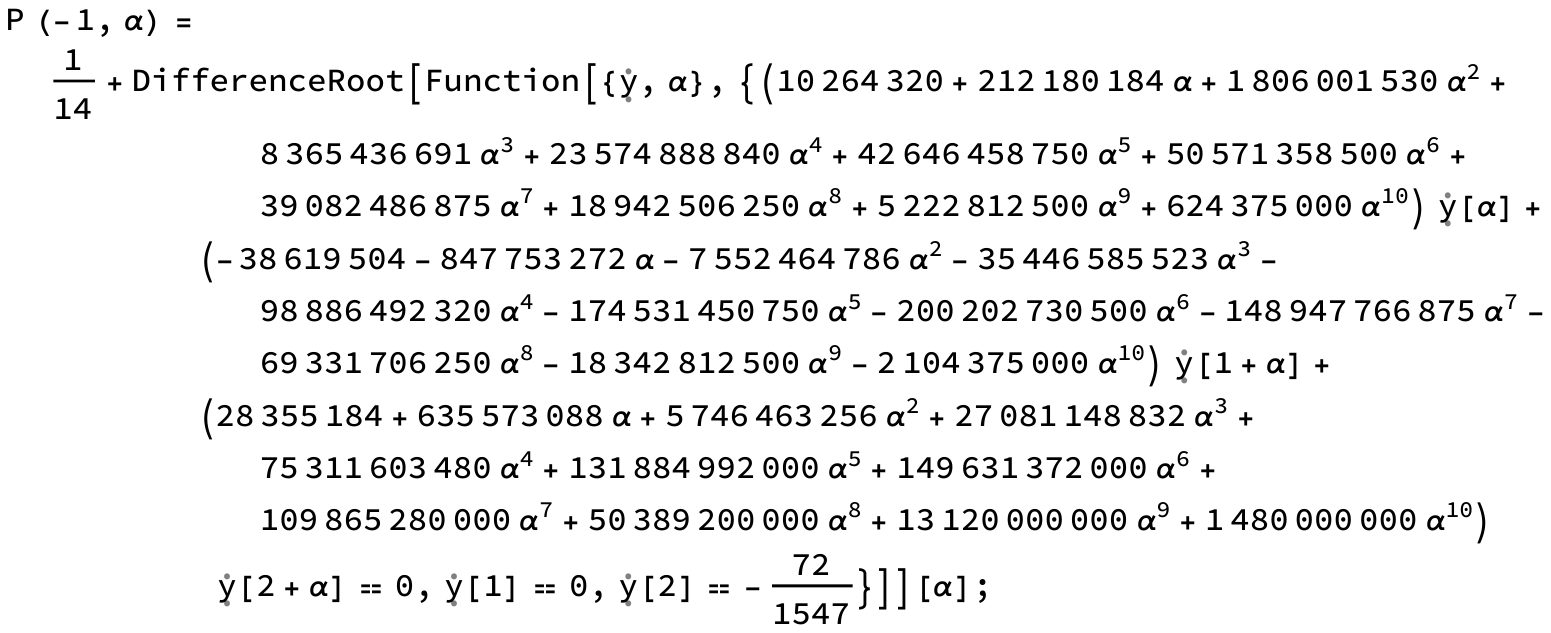}
\section{Further $P(f(\alpha),\alpha)$ formulas} \label{P(f(a),a)}
\includegraphics[page=1,scale=0.95]{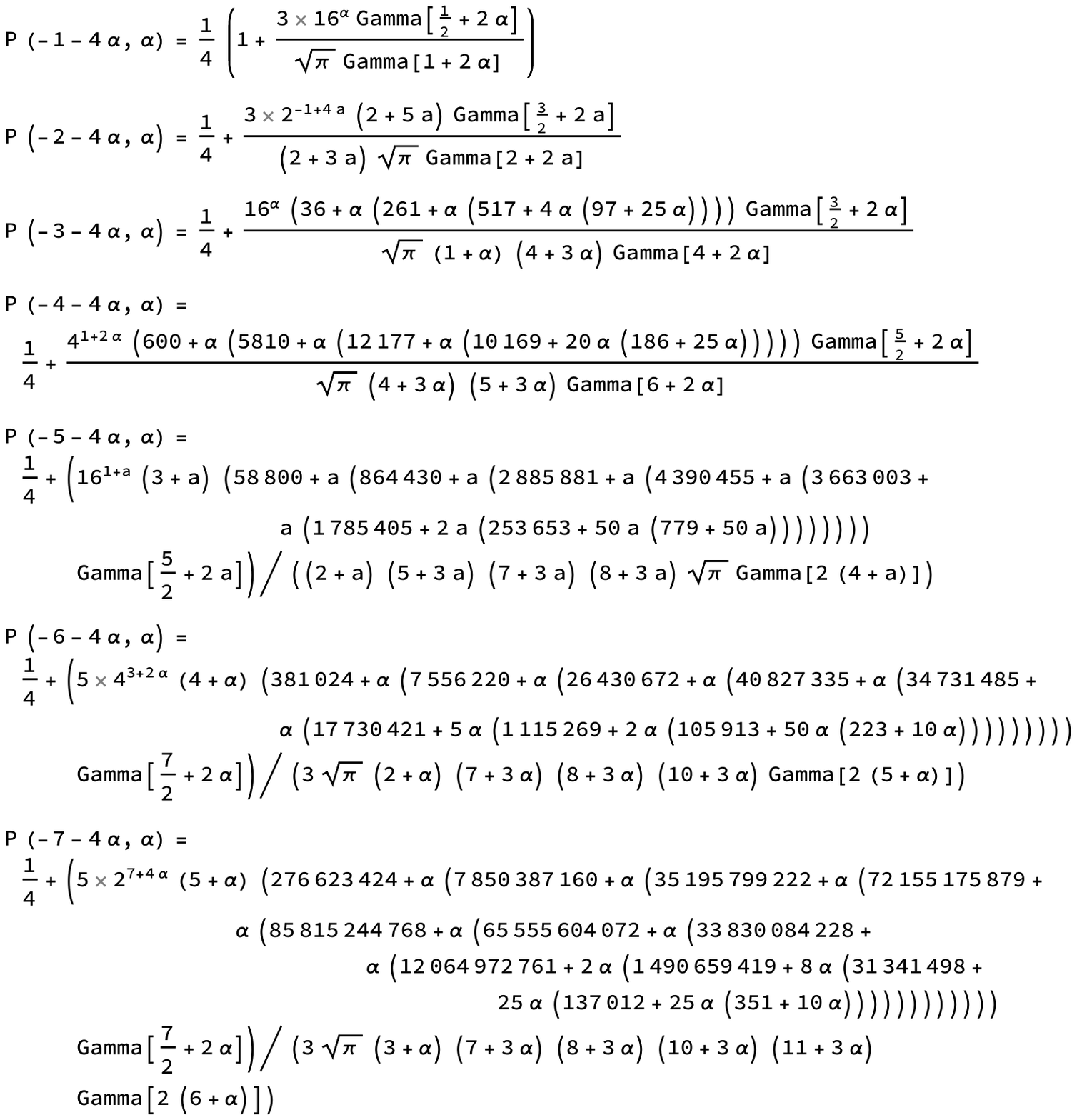}
\includegraphics[page=2,scale=0.95]{AdditionalPkaFormulas2.pdf}
\section{Formula for $P(-\alpha,\alpha)$} \label{fig:P(-a,a)}
\includegraphics[scale=0.95]{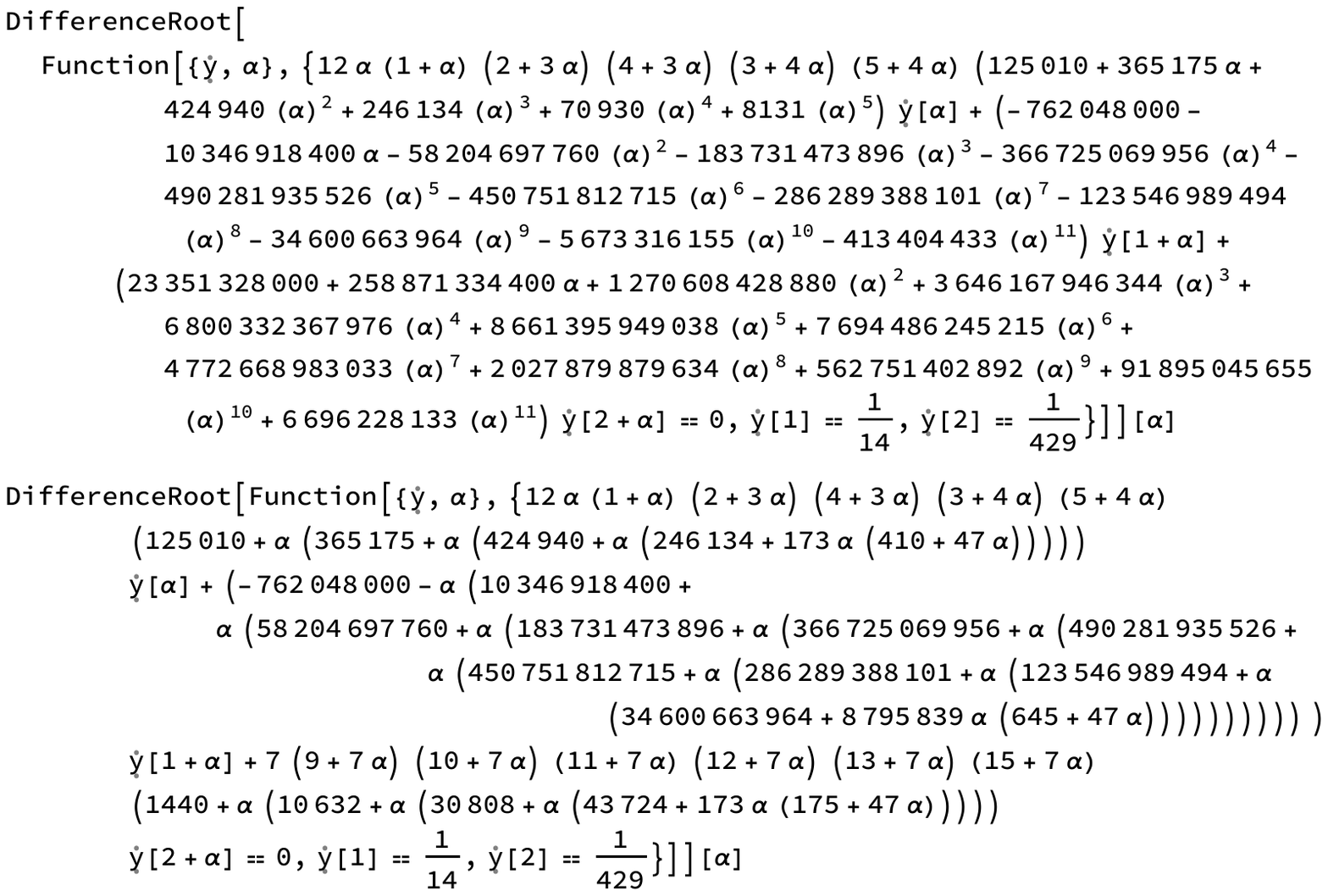}
\section{Formulas for leading coefficients of $p_{\alpha}(k)$} \label{CoefficientRules}
\includegraphics[page=1,scale=.9]{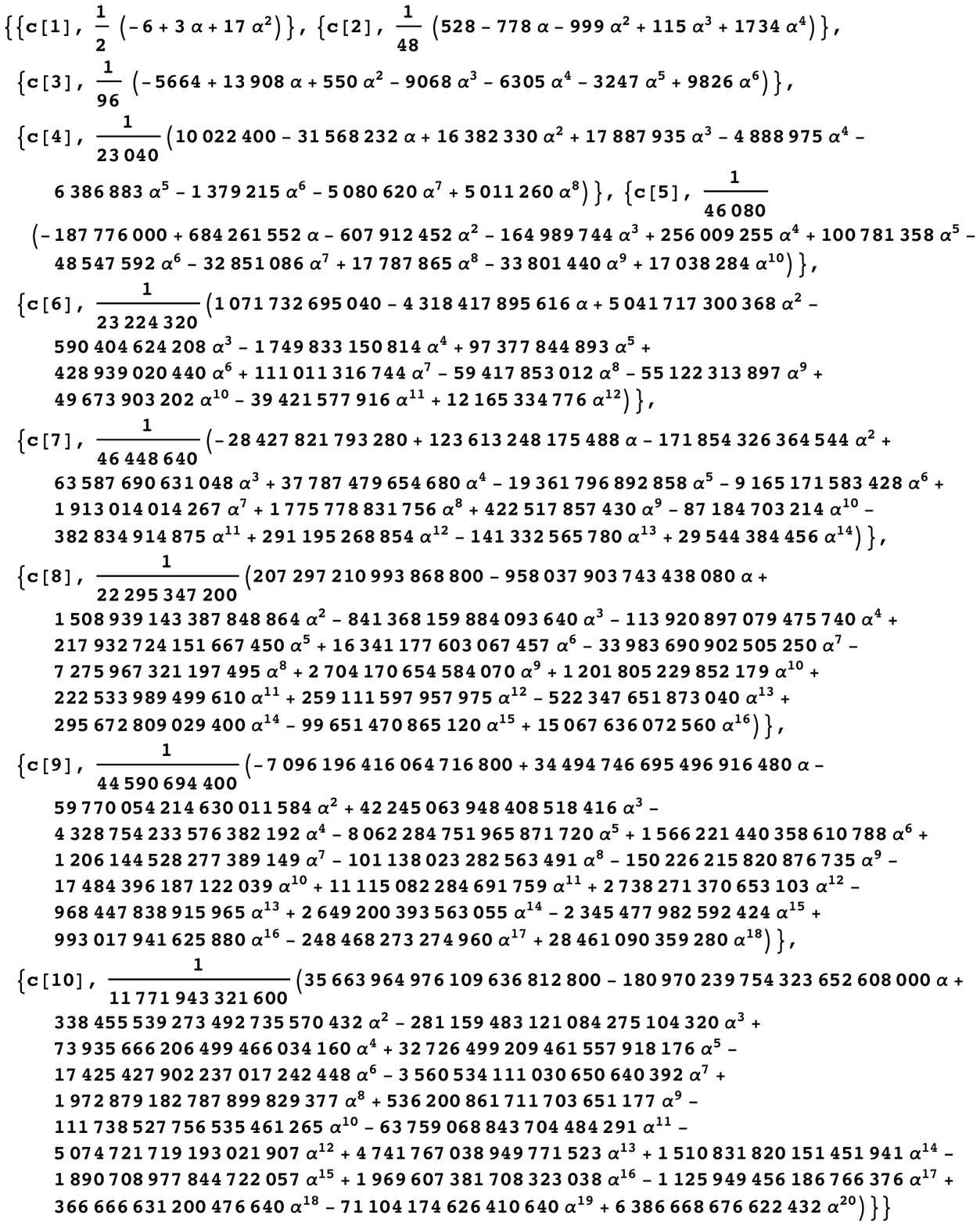}
\section{``Exterior'' separability probabilities}
\subsection{Inspheres}
The convex set of two-qubit states  possesses an ``insphere'' of maximum radius. The states within in it are {\it all} separable \cite{ZHSL,sbz}. So, one can ask what is the Hilbert-Schmidt separability probability {\it outside} of it, presuming the apparent {\it total} separability probability of $\frac{8}{33} \approx 0.242424$. Using the formulas in \cite{szHS}, we have 
$\frac{\pi ^6}{851350500}$ for the total volume of the two-qubit states, 
$\frac{1}{2 \sqrt{3}}$ for the radius of this insphere, and thus 
$\frac{\pi ^7}{567437270400 \sqrt{3}}$ for its 15-dimensional volume. This yields an exterior separability probability  of
\begin{equation}
E_{Insphere}^{two-qubits}=\frac{385 \sqrt{3} \pi -186624}{11 \left(35 \sqrt{3} \pi -69984\right)}  =
\frac{1}{1+\frac{1}{\frac{8}{25}-\frac{77 \pi }{38880 \sqrt{3}}}} \approx 0.240357.
\end{equation} 

Let us proceed similarly for the two-rebit states. We use, again, the pertinent formulas 
\cite[sec. 7]{szHS}, obtaining a total volume of $\frac{\pi ^4}{10080}$, a radius of the insphere of $\frac{1}{6 \sqrt{3}}$, and a 9-dimensional insphere volume of $\frac{\pi ^4}{24106163760 \sqrt{3}}$. This yields a separability probability 
(ever so slightly less than the presumed value of 
$ \frac{29}{64} \approx 0.453125000$) exterior to the insphere of
\begin{equation}
E_{Insphere}^{two-rebits}=\frac{128 \sqrt{3}-416118303}{64 \left(2 \sqrt{3}-14348907\right)} =
\frac{1}{1+\frac{1}{\frac{29}{35}-\frac{128}{167403915 \sqrt{3}}}} \approx 0.453124868.
\end{equation}
\subsection{Absolutely separable states}
Next, let us observe that these inspheres are themselves contained within the sets of {\it absolutely} separable states \cite{ver}--those states that can not be entangled through unitary transformations. In \cite[eq. (32)]{slater2009eigenvalues}, the result 
$\frac{6928 -2205 \pi}{16 \sqrt{2}} \approx 0.0348338$ was reported for the  two-rebit absolute separability probability. This leads to an exterior separability probability of 
\begin{equation}
E_{AbsSep}^{two-rebits}=\frac{29-13856 \sqrt{2}+4410 \sqrt{2} \pi }{2 \left(32-6928 \sqrt{2}+2205 \sqrt{2} \pi
   \right)} = \frac{1}{1+\frac{35}{29-13856 \sqrt{2}+4410 \sqrt{2} \pi }} \approx 0.433387744.
\end{equation}
Also, a considerably more complicated two-qubit formula 
\cite[eq. (34)]{slater2009eigenvalues} was given. The corresponding absolutely separable probability is approximately 0.00365826. This yields, proceeding similarly, to $E_{AbsSep}^{two-qubits} \approx 0.239643$.
\begin{acknowledgments}
I would like to express appreciation to 
Charles Dunkl for his many, many expert contributions and interactions in this research program in the past few years, and specifically for his important insights reported in sec.~\ref{DunklInsight}. Qing-Hu Hou has been very generous in his assistance also. Christioph Koutschan helped with the implementation of the fast Zeilberger algorithm in the  RISC package. Christian Krattenthaler provided advice at early stages of the research reported. Robert Israel, Brendan Godfrey and Michael Love have helpfully responded to a number of questions posted on the Mathematics and Mathematica Stack Exchanges.

\end{acknowledgments}

\bibliography{DifferenceHyper8}

\end{document}